\documentclass[
            aps,
            prd, superscriptaddress,
            nofootinbib,
            10pt]{revtex4-2}

    \usepackage[utf8]{inputenc}
    \usepackage[english]{babel}

    \usepackage{graphicx}
    
    \usepackage{booktabs}
    \usepackage{makecell}
    \usepackage{multirow}
    
    \usepackage{amsmath}
    \usepackage{amssymb}
    \usepackage{amsfonts}
    \usepackage{mathtools}
    \usepackage{bm}% bold math
    \usepackage{mathrsfs}
    \usepackage{upgreek}
    
    \usepackage{comment}
    \usepackage{xcolor}
    \usepackage[normalem]{ulem}
    
    \usepackage[hidelinks]{hyperref}
    
    \newcommand{\rmd}{\,\mathrm{d}} %derivative d
    \newcommand{\ii}{\mathrm{i}} %complex i
    \newcommand{\ee}{\mathrm{e}} %exponential
    
    \newcommand{\diag}{\mathrm{diag}}
    
    \newcommand{\p}{\partial}
    
    \newcommand{\ket}[1]{\left| #1 \right\rangle}
    
    \newcommand{\braket}[2]{\left\langle #1 \middle| #2 \right\rangle}

    \DeclareMathOperator{\arcsinh}{arcsinh}
    
    \newcommand{\rref}[1]{(\ref{#1})}
    \newcommand{\figref}[1]{Fig.~\ref{#1}}
    
    \usepackage{calligra}
    \DeclareMathAlphabet{\mathcalligra}{T1}{calligra}{m}{n}
    \DeclareFontShape{T1}{calligra}{m}{n}{<->s*[2.2]callig15}{} % Scales the font appropriately
    
    \newcommand{\scriptr}{\mathcalligra{r}\,}

\begin{document}

    \title{Neutrino Oscillation in a Static Dihole Geometry}
    
    \date{\today}

    \author{Rog\'erio Capobianco}
        \email{rogerio.capobianco@gmail.com}
        \affiliation{Universidade do Estado de Santa Catarina, Joinville, 89219-710, SC, Brazil}
        \affiliation{Hanse-Wissenschaftskolleg, Lehmkuhlenbusch 4, 27753 Delmenhorst, Germany}
    
    \author{Rafael C. R. de Lima}
        \email{rafael.lima@udesc.br}
        \affiliation{Universidade do Estado de Santa Catarina, Joinville, 89219-710, SC, Brazil}

    \begin{abstract}
       We investigate neutrino flavor oscillations in the static equal-mass Majumdar-Papapetrou dihole geometry, restricting the propagation to the symmetry plane. Within the null approximation, we derive the leading contribution to the neutrino phase and numerically evaluate the geodesic integrals. For two neutrino flavors, interference between different trajectories introduces a dependence on the individual neutrino masses and on the neutrino mass ordering for nonmaximal mixing. We analyze the transition probability in different spacetime configurations possessing two light rings, a merged light ring (critical configuration), and no light rings in the equatorial plane. In the weak-lensing regime, the transition probability generally increases with the black holes' individual masses, while variations in the half-separation produce no qualitative change in the transition probability. In stronger-lensing configurations, variations in both the mass and half-separation modify the heights and positions of the extrema, with no monotonic increase or decrease in the transition probability. For the critical and no-light-ring configurations studied, we identify three zero-winding trajectories connecting the source and detector, including one passing near the symmetry axis. The probability remains oscillatory, and its oscillatory pattern depends strongly on the neutrino energy.
    \end{abstract}

    \maketitle

    %%%%%%%%%%%%%%%%%%%%%%%%%%%%%%%%%%%%%%%%%%%%%%%%%%%%%%%%%%%%
    \section{Introduction}
    %%%%%%%%%%%%%%%%%%%%%%%%%%%%%%%%%%%%%%%%%%%%%%%%%%%%%%%%%%%%
    
    Almost a century after their first theoretical prediction, neutrinos, the lightest known fermions to this day, still present many mysteries and challenges for our understanding of nature. Originally proposed in the context of beta decay \cite{Bilenky:2012qb}, their existence was then established through detection of antineutrinos \cite{Cowan:1956rrn}. Subsequently, observations of solar and atmospheric neutrinos provided decisive evidence for neutrino flavor transformations \cite{Super-Kamiokande:1998kpq,SNO:2002tuh}, establishing that neutrinos have non-zero masses and therefore requiring an extension of the minimal Standard Model of particle physics. Neutrino flavor states are understood as coherent superpositions of mass eigenstates, whose relative phases evolve during propagation, giving rise to oscillations. Experiments have constrained the mixing parameters and mass-squared differences, but the absolute neutrino mass scale and the mass hierarchy are still open questions \cite{ParticleDataGroup:2026mpi}. In flat spacetimes, the standard vacuum treatment of oscillation probability along a single path depends on mass-squared differences rather than each particle's individual mass. In curved geometries, however, gravitational lensing can provide multiple propagation paths whose contributions can interfere coherently for the quantum state. Interference between these paths can introduce dependence on the individual neutrino masses and the mass ordering \cite{Cardall:1996cd,Fornengo:1996ef,Swami:2020qdi}. Neutrinos are produced in a variety of astrophysical processes, including nuclear reactions in stellar interiors and core-collapse supernovae, while a relic neutrino background is expected from the early universe. These neutrino populations carry information about the physical conditions in their astrophysical sources as well as the thermal history of the universe \cite{Vitagliano:2019yzm, Lesgourgues:2006nd, Hannestad:2006zg, Muller:2019upo}. Their propagation and the gravitational effects on their quantum state therefore motivate the study of neutrinos in strong gravitational fields.       
    
    General relativity (GR) is the current classical theory of gravity. Whereas Newtonian gravity describes an attractive force acting in absolute space and time between masses, in Einstein's formulation, gravity is an effect of the spacetime curvature, relating gravitational interaction and spacetime geometry. Its predictions have been remarkably successful in experimental and observational tests, from Solar System measurements to binary pulsars \cite{Will:2014kxa}. Recently, the direct detection of gravitational waves from merging black holes and neutron stars \cite{LIGO_binary1,LIGO_binary2,LIGO_binary3,LIGO_binary4} and the horizon-scale imaging of black holes M87* and Sagittarius A* \cite{EventHorizonTelescope:2022wkp,EventHorizonTelescope:2019dse} have opened new possibilities for exploring gravity in the strong-field regime. A fundamental approach to understanding physical phenomena in curved spacetime is the description of geodesic motion. Within GR, freely falling test particles follow timelike or null geodesics, depending on whether they are massive or massless, revealing how spacetime curvature affects their propagation. Null geodesics govern gravitational lensing and the formation of black hole shadows. In the presence of strong gravitational fields around compact objects, circular null orbits, known as \textit{light rings}, can exist \cite{Cunha:2017qtt}. For massive particles, the existence and stability of circular timelike geodesics determine characteristic orbital scales, including the innermost stable circular orbit, which plays an important role in accretion-disk models \cite{abramowicz1978relativistic,Abramowicz:2011xu}. The understanding of those trajectories also provides the geometrical basis for evaluating the phases accumulated by neutrinos propagating in curved-geometry backgrounds.   
    
    Describing binary black hole systems within the context of GR poses considerable difficulties. On the one hand, the dynamical evolution of such systems, particularly during merger, requires computationally demanding numerical-relativity techniques. On the other hand, the use of analytical approaches and exact solutions can often shed light on the physics around those objects. However, despite the high interest, a general exact solution describing an astrophysically dynamical binary system is not known. There is an equilibrium situation between two sources that can be obtained even in Newtonian gravity -- in this case, the electrostatic repulsion is set to counterbalance the gravitational attraction, allowing a collection of charged particles with suitable charge-to-mass ratios to reach a state of mechanical equilibrium. The exact balance between gravitational attraction and electrostatic repulsion still holds in general relativity -- the solution that describes this physical scenario is known as the Majumdar-Papapetrou (MP) multi-black hole solution \cite{Majumdar_MPsol,Papapetrou_MPsol}. Despite its early derivation, the interpretation of the MP spacetime as a multi-centered solution of any number of extremal Reissner-Nordstr\"om black holes came years later \cite{Hartle_MPsol}. Remarkably, this is an exact solution of the coupled Einstein-Maxwell equations, static, asymptotically flat, and free of any conical singularities. For the two-source case, the solution is, in addition, axially symmetric. Despite such nice characteristics, the MP solution exhibits rich and complex physics; the geodesic equations are generally non-separable, and both massless and massive particles can exhibit chaotic motion \cite{chandrasekhar1989two,contopoulos1990periodic,contopoulos1991periodic,Dettmann:1994dj,Dettmann:1995ex}. However, there is a special scenario: when the MP is considered to represent a spacetime containing two equal-mass black holes, then the symmetry plane halfway between the two black holes is invariant under geodesic motion. Furthermore, the static multi-centered black holes as described by the MP solution have drawn such attention that the solution has been further extended to higher dimensions \cite{Lemos:2005md,Lemos:2006sj}, and generalized to include a positive cosmological constant  \cite{Kastor:1992nn,Kastor:1993mj,Nakao:1994mm}. A detailed discussion of gravitational lensing in the symmetry plane of the dihole MP spacetime can be found in \cite{MP_lensing_complete}. Although its extremal charges and static equilibrium distinguish the MP solution from astrophysical binaries, it provides an exact setting for investigating physical effects in a two-center gravitational field. The dihole geometry has already been used to investigate binary BH shadows and chaotic scattering, stable photon trapping, and the influence of a companion on black hole geometry and curvature \cite{Shipley:2016omi,Dolan:2016bxj,Semerak:2016gfz}.
    
    In this paper, we focus on neutrino oscillations in the equal-mass dihole Majumdar-Papapetrou geometry, restricting the motion to the symmetry plane. We adopt here the null approximation, in which the propagation path for relativistic neutrinos is approximated by a corresponding null geodesic while retaining the leading mass-dependent contribution to the oscillation phase. The formalism to explore neutrino oscillations in curved spacetime has established covariant treatments of the propagation phase and explored how gravitational fields modify flavor evolution as neutrinos propagate across curved geometries \cite{Cardall:1996cd,Fornengo:1996ef}. Gravitational lensing further allows for interference between contributions from distinct paths connecting the source and detector, producing an interplay between flavor oscillation and path interference \cite{Crocker:2003cw}. In contrast to standard single-path vacuum oscillations, within a coherent multi-path formulation, these interference patterns can lead to dependence on individual neutrino masses and sensitivity to mass ordering \cite{Swami:2020qdi}. Furthermore, wave-packet treatments have also examined the decoherence in neutrino lensing, showing that the suppression of interference depends on the neutrino's individual masses, propagation trajectories and wave-packet widths \cite{Swami:2021wbf}. Consequently, exploring neutrino oscillations in curved spacetime provides a framework for investigating neutrino properties and their interplay with gravity. These effects were already explored for static black hole spacetimes, including Schwarzschild \cite{Cardall:1996cd,Fornengo:1996ef} and related flavor interferometry in Reissner-Nordström backgrounds \cite{Alexandre:2025qip}. Attempts to include rotation have also been explored, in the cases of Kerr and Kerr-Newman backgrounds, including analyses for special cases and in the weak-field regime \cite{Ren:2010yf,Li:2026olr,Swami:2022xet}. Furthermore, neutrino oscillations have also been explored in alternative gravitational formulations and parametrized black hole geometries \cite{Chakrabarty:2023kld,Wang:2025nyl,Alloqulov:2024sns,MohseniSadjadi:2026fkh,AraujoFilho:2025rzh}.
    
    This paper is organized as follows. In Sec.~\ref{sec:framework}, we introduce the general framework to study neutrino oscillations in curved spacetimes, deriving the covariant phase in the eikonal approximation and introducing the flavor probability for a multi-path formalism. In Sec.~\ref{sec:dihole}, we introduce the Majumdar-Papapetrou solution, specializing to the equal-mass dihole solution and obtaining the phase at leading order within the null approximation; we close by discussing the purely radial motion. In Sec.~\ref{sec:lensing}, we discuss gravitational lensing of neutrinos in detail, including the numerically obtained transition probability for MP configurations with two LRs, a merged LR, and no LRs. Finally, Sec.~\ref{sec:conclusion} summarizes our findings and discusses possible extensions. Throughout the theoretical foundations, we adopt units in which \(G=c=\hbar=1\); physical units are recovered for the numerical analysis. We adopt \( (+,-,-,-) \) for the metric signature and the Einstein summation convention for repeated spacetime and local Lorentz indices. Sums over neutrino mass eigenstates and propagation paths are written explicitly.  
    
    %
    %
    %
    %
    %
    %
    %
    %
    %
    %
    %%%%%%%%%%%%%%%%%%%%%%%%%%%%%%%%%%%%%%%%%%%%%%%%%%%%%%%%%%%%
    \section{Neutrino oscillations in curved spacetime}\label{sec:framework}
    %%%%%%%%%%%%%%%%%%%%%%%%%%%%%%%%%%%%%%%%%%%%%%%%%%%%%%%%%%%%
    
    Before specializing to the dihole geometry, we outline how the oscillation phase emerges from the Dirac equation in a curved background. Besides fixing the conventions used throughout, this makes explicit the approximations entering the calculation and shows in which sense the familiar flat-spacetime formula is recovered as a particular case.
    
    \subsection{The Dirac equation in the tetrad formalism}
    
    Spinor fields do not carry a representation of the group of general coordinate transformations, as tensors do, so the Dirac equation in a curved spacetime is formulated with respect to a local Lorentz frame. Such a frame is provided by a tetrad field $e^{\hat a}_{\ \mu}(x)$, defined by
    
    \begin{equation}\label{tetrad}
        g_{\mu\nu} = \eta_{\hat a \hat b} \, e^{\hat a}_{\ \mu} \, e^{\hat b}_{\ \nu} \, , \qquad \eta_{\hat a \hat b} = \diag(1,-1,-1,-1) \, ,
    \end{equation}
    
    where Greek indices refer to the spacetime coordinates and hatted Latin indices to the local Minkowski frame, the inverse tetrad satisfying $e^{\hat a}_{\ \mu} \, e_{\hat b}^{\ \mu} = \delta^{\hat a}_{\ \hat b}$. Equation \rref{tetrad} fixes the tetrad only up to a local Lorentz transformation, which encodes the freedom of choosing the orientation and the state of motion of the observer at each spacetime point. The position-dependent Dirac matrices are then built as
    
    \begin{equation}\label{gamma_curved}
        \gamma^\mu(x) = e_{\hat a}^{\ \mu}(x) \, \gamma^{\hat a} \, , \qquad \left\{ \gamma^\mu , \gamma^\nu \right\} = 2 \, g^{\mu\nu} \, ,
    \end{equation}
    
    with $\gamma^{\hat a}$ the constant Dirac matrices of flat spacetime. Each mass eigenstate is associated with a spinor field $\psi_k$ obeying the curved Dirac equation:
    
    \begin{equation}\label{dirac_curved}
        \left[ \ii \, \gamma^\mu(x) \, D_\mu - m_k \right] \psi_k = 0 \, ,
    \end{equation}
    
    where the spinorial covariant derivative and the spin connection read \cite{Cardall:1996cd}
    
    \begin{equation}\label{spin_connection}
        D_\mu = \p_\mu + \Gamma_\mu \, , \qquad \Gamma_\mu = \frac{1}{8} \, \omega_{\mu \hat a \hat b} \, \left[ \gamma^{\hat a} , \gamma^{\hat b} \right] \, ,
    \end{equation}
    
    with $\omega_{\mu \hat a \hat b} = e_{\hat a}^{\ \nu} \, \nabla_\mu e_{\hat b \nu}$ the Ricci rotation coefficients.
    
    About \rref{dirac_curved}, we emphasize -- it is the mass eigenstates, and not the flavor states, that obey \rref{dirac_curved}. A flavor field is the coherent superposition $\psi_\alpha = \sum_k \mathcal{U}^*_{\alpha k} \, \psi_k$ and consequently does not propagate with a definite mass, which is the field-theoretical statement of the mixing.

    \subsection{Eikonal limit and the covariant phase}
    
    Neutrinos of astrophysical interest are extremely short-wavelength probes of the geometry: for an energy $E \sim 10 \, \mathrm{MeV}$ and a curvature scale set by a solar-mass compact object, the ratio between the de Broglie wavelength and the curvature radius is of order $10^{-17}$. We may therefore solve \rref{dirac_curved} in the eikonal (WKB) approximation, writing
    
    \begin{equation}\label{wkb_ansatz}
        \psi_k(x) = \left[ \varphi_k(x) + \mathcal{O}\left( \lambda_{\rm dB} \right) \right] \, \ee^{- \ii \, \Phi^{(k,p)}(x)} \, ,
    \end{equation}
    
    where $\Phi^{(k,p)}$ is a rapidly varying real phase and $\varphi_k$ a slowly varying spinor amplitude. Note that here we associate two indices with the phase: $(k)$ labels the neutrino mass eigenstate, whereas $(p)$ labels the propagation path, corresponding to a particular branch of the eikonal solution that connects the source and the detector. The introduction of the latter is of particular interest since, in gravitational-lensing settings, multiple paths can connect the source and the detector. In the null approximation we adopt here, the geometrical trajectory associated with each branch can be determined independently of the neutrino mass at leading order in the approximation, while the neutrino mass is still retained during the propagation phase.
    Inserting \rref{wkb_ansatz} into \rref{dirac_curved} and retaining the leading contribution, the one in which the derivative acts on the phase, yields the algebraic relation
    
    \begin{equation}\label{leading_eikonal}
        \left[ \gamma^\mu \, \p_\mu \Phi^{(k,p)} - m_k \right] \varphi_k = 0 \, ,
    \end{equation}
    
    in which the spin connection is absent at this order. Multiplying \rref{leading_eikonal} from the left by $\left[ \gamma^\nu \, \p_\nu \Phi^{(k,p)} + m_k \right]$ and using the Clifford algebra \rref{gamma_curved}, a non-trivial amplitude requires
    
    \begin{equation}\label{hamilton_jacobi}
        g^{\mu\nu} \, \p_\mu \Phi^{(k,p)} \, \p_\nu \Phi^{(k,p)} = m_k^{\, 2} \, ,
    \end{equation}
    
    which is the relativistic Hamilton--Jacobi equation. The identification
    
    \begin{equation}\label{momentum_gradient}
        p^{(k,p)}_\mu = \p_\mu \Phi^{(k,p)}
    \end{equation}
    
    therefore reproduces the mass-shell condition where $p^{(k,p)}$ is the four-momentum along a geodesic $\gamma_p$. 
    Moreover, since $p^{(k,p)}_\mu$ is a gradient, $\nabla_\mu p^{(k,p)}_\nu = \nabla_\nu p^{(k,p)}_\mu$, and \rref{hamilton_jacobi} gives at once
    
    \begin{equation}\label{geodesic_from_HJ}
        p^{(k,p) \, \nu} \, \nabla_\nu \, p^{(k,p)}_\mu = \frac{1}{2} \, \nabla_\mu \left( p^{(k,p) \, \nu} \, p^{(k,p)}_\nu \right) = \frac{1}{2} \, \nabla_\mu m_k^{\, 2} = 0 \, .
    \end{equation}
    
    The integral curves of $p^{(k,p)}_\mu$ are thus geodesics, and integrating \rref{momentum_gradient} along the \(p\)-th geodesic-path $\gamma_p$ connecting the production to the detection event yields the covariant phase \cite{Cardall:1996cd,Fornengo:1996ef}
    
    \begin{equation}\label{covariant_phase_wkb}
        \Phi^{(k,p)} = \int_{\gamma_p} p^{(k,p)}_\mu \, \rmd x^\mu \, ,
    \end{equation}
    
    where $\gamma_p$ represents the trajectory the neutrino propagates along from the source to the detector. The observable interference is then governed by phase differences involving different masses (k), paths (p), or both. These differences constitute the central quantities to be evaluated for the MP dihole geometry.
    
    Two comments are in order. The spin connection, absent from \rref{hamilton_jacobi}, enters only at the following order of the eikonal expansion, where it governs the propagation of the amplitude $\varphi_k$ along the trajectory. Its effect is a rotation of the neutrino polarization, the gravitational analogue of a spin precession, rather than a contribution to the dynamical phase responsible for the flavor oscillation \cite{Cardall:1996cd}. Furthermore, the mass appears explicitly in \rref{hamilton_jacobi}, so that each mass eigenstate is strictly associated with a distinct timelike geodesic. This is the root of the ambiguity discussed in Sec.~\ref{sec:dihole}: the null approach amounts to evaluating the trajectory along the corresponding null geodesic while retaining the mass in the momentum. Hence, for ultrarelativistic neutrinos, the trajectory is evaluated at zeroth order -- this is the respective null geodesic -- while the leading-mass order is retained in the propagation phase. As a consequence, the propagation path can be treated independently of the neutrino mass; thus the phase difference for a fixed path depends only on the squared masses. 
    
    \subsection{Oscillation probability}
    
    Since all mass eigenstates contribute coherently at the detection event, the flavor state at the detector is a simple superposition of the mass eigenstates with their corresponding propagation factors,
    
    \begin{equation}\label{flavor_at_detector}
        \ket{\nu_\alpha (x_D)} = \mathcal{N} \, \sum_k \mathcal{U}^*_{\alpha k} \, \mathcal{P}_k
        % \ee^{- \ii  \Phi^{(k, p)}} 
        \ket{\nu_k} \, ,
    \end{equation}
    
    where $\mathcal{N}$ is a normalization constant
    
    \begin{equation}\label{normalization}
        \lvert \mathcal{N} \rvert^2 = \left( \sum_{k} \lvert \mathcal{U}_{\alpha k} \rvert^2 \lvert \mathcal{P
        }_k \rvert^2 \right)^{-1} \, .
    \end{equation}
    
    The propagation factor $\mathcal{P}_k$ accounts for all paths available to the \(k\)-th mass eigenstate. Thus, at the detector it is 
    
    \begin{equation}\label{Path_shortcut}
        \mathcal{P}_k = \sum_{p=1}^{N_p} \ee^{-\ii \Phi^{(k,p)}} \, ,
    \end{equation}
    
    which represents the coherent sum of the propagating amplitudes associated with the \(k\)-th mass eigenstate over the $N_p$ considered paths. Hence, the probability that a neutrino produced with the flavor $\alpha$ is detected with the flavor $\beta$ is
    
    \begin{equation}\label{oscillation_probability}
        P_{\alpha \to \beta} = \left| \braket{\nu_\beta}{\nu_\alpha(x_D)} \right|^2 
        = \lvert \mathcal{N} \rvert^2  \sum_{k,j} \mathcal{U}_{\beta k} \, \mathcal{U}^*_{\alpha k} \, \mathcal{U}^*_{\beta j} \, \mathcal{U}_{\alpha j} \, \mathcal{P}_k \mathcal{P}^{*}_j
        % \ee^{- \ii \, \Delta \Phi_{kj} } \, .
    \end{equation}
    
    Note that when only a single propagation path, say $p=1$, contributes, the whole dependence on the geometry is carried by the relative phase $ \Delta \Phi^{(k,j)} =  \Phi^{(k,p=1)} - \Phi^{(j,p=1)}$. However, in general, we keep the relation 
    
    \begin{equation}\label{path_contribution_product}
        \mathcal{P}_k \mathcal{P}_j^{*} = \sum_{p,q} \ee^{-\ii \left( \Phi^{(k,p)} - \Phi^{(j,q)} \right) } \, .
    \end{equation}
    
    It is instructive to recover the flat-spacetime limit. In Minkowski spacetime, the four-momentum is constant along the trajectory. A unique path connects the source and detector for a freely propagating neutrino, which is a straight line of length $L$. Hence, $\Phi^{(k)} = E_k \, T - \lvert \vec{p}_k \rvert \, L$, where $T$ is the propagation time. For ultrarelativistic neutrinos produced with a common energy $E$ one has $\lvert \vec{p}_k \rvert \simeq E - m_k^{\, 2}/2E$ and at leading order $T \simeq L$, whence
    
    \begin{equation}\label{flat_phase}
        \Phi^{(k)} \simeq \frac{m_k^{\, 2}}{2E} \, L \, ,
    \end{equation}
    
    and \rref{oscillation_probability} reduces, for two flavors mixed by an angle $\theta$, to the familiar expression
    
    \begin{equation}\label{two_flavor}
        P_{\alpha \to \beta} = \sin^2 2\theta \; \sin^2 \left( \frac{\Delta m^2 \, L}{4 E} \right) \, , \qquad \alpha \neq \beta \, ,
    \end{equation}
    
    with $\Delta m^2 = m_2^{\, 2} - m_1^{\, 2}$ and oscillation length $L_{\rm osc} = 4 \pi E / \lvert \Delta m ^2 \rvert$. These approximations are consistent with the null formalism employed in the following section.
    
    In a curved background, the structure of \rref{oscillation_probability} is unchanged: for a single path within the present treatment, the entire effect of the geometry is encoded in the scalar $\Delta \Phi^{(k,j)}$, evaluated along the geodesic joining source and detector. In general gravitational-lensing settings, multiple null geodesics may connect the same source and detector. Therefore, neutrino oscillation in a curved spacetime involves not just the usual interference between different mass eigenstates propagating along the same path, but also between same-mass eigenstates propagating along different paths, as well as between different mass eigenstates propagating along different paths, provided that the corresponding amplitudes remain coherent from the production event to the detection event. In the next section, this formalism will be applied to describe neutrino oscillation in the MP dihole geometry.

    %%%%%%%%%%%%%%%%%%%%%%%%%%%%%%%%%%%%%%%%%%%%%%%%%%%%%%%%%%%%
    \section{Neutrino oscillation in a static dihole solution}\label{sec:dihole}
    %%%%%%%%%%%%%%%%%%%%%%%%%%%%%%%%%%%%%%%%%%%%%%%%%%%%%%%%%%%%
    
    The Majumdar-Papapetrou spacetime is an exact solution of the Einstein-Maxwell field equations describing a collection of extremal Reissner-Nordstr\"om black holes in mutual equilibrium. The solution represents an exact balance between the gravitational attraction and electrostatic repulsion that allows the black holes to remain at rest for an arbitrary number of black holes. In Cartesian coordinates, the solution is represented by the line element
    
    \begin{equation}\label{MP_Cartesian}
        \rmd s^2 = U^{-2} \rmd t^2 - U^2 \left( \rmd x^2 + \rmd y^2 + \rmd z^2 \right) \, ,
    \end{equation}
    
    with 
    
    \begin{equation}
        U = 1 + \sum_{i} \frac{M_i}{ R_i } \, , \ \ \qquad \ \ \ R_i^{\,2} = (x-x_i)^2 + (y-y_i)^2 + (z-z_i)^2 \, . 
    \end{equation}
    
    Each source has mass $M_i$, charge $\lvert Q_i \rvert = M_i$, and is located at $(x_i,y_i,z_i)$. Furthermore, the associated electromagnetic potential is $A = U^{-1} \, \rmd t$. In general, with the ansatz \rref{MP_Cartesian}, the Einstein-Maxwell field equations reduce to the Laplace equation for $U$, allowing multicenter solutions to be constructed by linear superposition. The coordinates introduced above cover the region exterior to the horizons of each source. Although the metric appears singular at $R_i=0$, these points correspond to the degenerate event horizons of the sources rather than curvature singularities; thus, in such coordinates a horizon of finite area is mapped to a single point \cite{Hartle_MPsol}. The spacetime admits an analytic extension across each horizon \cite{Carter_MPsol}.   
    
    Throughout this work, we focus on the equal-mass dihole configuration, placing each black hole along the symmetry axis at $z=\pm a$, such that their coordinate separation is $2a$. In isotropic spherical coordinates, the metric becomes
    
    \begin{equation}\label{MP_metric}
        \rmd s^2 = U^{-2} \rmd t^2 - U^2\left( \rmd r^2 + r^2 \rmd \vartheta^2 + r^2 \sin^2\vartheta \rmd \varphi^2  \right) 
    \end{equation}
    
    where
    
    \begin{equation}
        U = 1 + \frac{M}{\sqrt{r^2 + a^2 - 2ar \cos \vartheta }} + \frac{M}{\sqrt{r^2 + a^2 + 2ar \cos \vartheta }} \, . 
    \end{equation}
    
    Note that the spacetime ADM mass is $2M$. The solution is static and axially symmetric, which means it admits two Killing vector fields $\partial_t$ and $\partial_\varphi$, and these isometries are also reflected in the geodesic motion, leading to two conserved quantities: the energy and azimuthal angular momentum. Moreover, the dihole solution is asymptotically flat.
    
    Despite the relatively simple form of the metric \rref{MP_metric}, its geometrical properties are quite complicated. The spacetime is algebraically general (Petrov type I), while the geodesic equations cannot, in general, be separated for either massless or massive particles. The resulting geodesic dynamics exhibits a rich structure. Null geodesics can either be captured by one of the two black holes or escape from the binary system, with unstable periodic orbits playing a fundamental role in distinguishing the outcomes \cite{chandrasekhar1989two,contopoulos1990periodic}. Timelike geodesics can additionally exhibit regular bound motion associated with stable periodic orbits \cite{contopoulos1991periodic}. Due to the lack of complete integrability, chaotic motion can also arise for both massive and massless particles \cite{Dettmann:1994dj,Dettmann:1995ex}. In the present work, we are mainly interested in the escaping null geodesics connecting the source and detector, which constitute the relevant trajectories for the gravitational-lensing configuration considered here. A detailed description of the gravitational lensing in the dihole geometry was presented in \cite{MP_lensing_complete}. 
    
    %%%%%%%%%%%%%%%%%%%%%%%%%%%%%%%%%%%%%%%%%%%%%%%%%%%%%%%%%%%%%%%%%%%%%%%%%%%%%%%
    \subsection{General Framework}
    %%%%%%%%%%%%%%%%%%%%%%%%%%%%%%%%%%%%%%%%%%%%%%%%%%%%%%%%%%%%%%%%%%%%%%%%%%%%%%%
    
\begin{figure}[h!]
        \centering
        \includegraphics[width=0.5\linewidth]{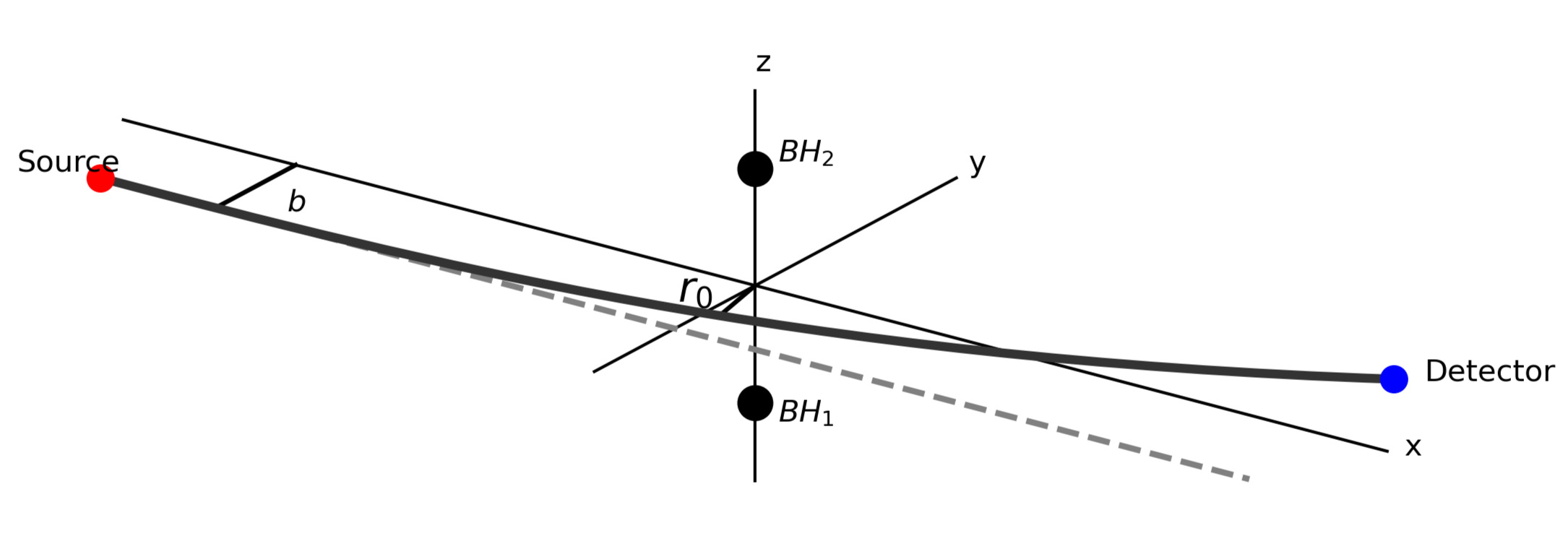}
        \caption{Schematic representation of the gravitational lensing generated by the MP dihole solution. A particle produced at the source is deflected due to the presence of the two BHs until it reaches the detector. Here, $b$ is the impact parameter and $r_0$ is the radius of closest approach. }
        \label{fig:neutrino_lens_schematic}
    \end{figure}

    We consider a neutral particle moving along a geodesic that bends due to the spacetime curvature. The motion is restricted to be in the symmetry plane between the two BHs, that is, $\vartheta = \pi/2$, which is an invariant plane of the geodesic motion. A schematic representation is presented in \figref{fig:neutrino_lens_schematic}. In that case, the function $U$ takes a particularly simple form: 
    
    \begin{equation}\label{U_eq_plane}
        U = 1 + \frac{2M}{\sqrt{r^2+a^2}} \, .
    \end{equation}
    
    For a particle moving along a geodesic path, labeled as $p$, in a curved spacetime, the four-momentum is 
    
    \begin{equation}
        p^{(k,p)}_\mu = m_k \, g_{\mu\nu} \, \frac{\rmd x^\nu}{\rmd \lambda} \, ,
    \end{equation}
    
    where $\lambda$ is the proper time. The reference null geodesic is parametrized by an affine parameter. Since in the following calculations we consider only a single generic geodesic path, the label $p$ will temporarily be suppressed and restored when different trajectories are considered.   
    
    For the MP solution \rref{MP_metric} on the equatorial plane, $p^{(k)}_\vartheta = 0$ and the non-vanishing components are: 
    
    \begin{eqnarray}\label{canonical_momenta}
        p^{(k)}_t &=& \frac{m_k}{U^2} \frac{\rmd t}{\rmd \lambda} \, ,  \label{canonical_t}  \\
        p^{(k)}_\varphi &=& - \, m_k \, r^2 U^2 \, \frac{\rmd \varphi}{\rmd \lambda} \, , \label{canonical_varphi} \\
        p^{(k)}_r &=& - \, m_k \, U^2 \frac{\rmd r}{\rmd \lambda} \, . \label{canonical_r} 
    \end{eqnarray}
    
    Here, recall the spacetime isometries. Since the MP solution is static and axially symmetric, it admits two Killing vector fields $\p_t$ and $\p_\varphi$. Consequently, the coordinates $t$ and $\varphi$ are cyclic variables. Hence, there are two conserved quantities along a geodesic: the particle's energy and azimuthal angular momentum, 
    
    \begin{equation}\label{cycvar}
        p^{(k)}_t = E_k \, , \ \ \ \qquad \ \ \ p^{(k)}_\varphi = -l_k \, ,
    \end{equation}
    
    with $E_k > 0$.
    Solving \rref{canonical_t} and \rref{canonical_varphi} together with the conserved quantities \rref{cycvar} for the velocity fields yields:
    
    \begin{eqnarray}
        m_k \, \frac{\rmd t}{\rmd \lambda} &=& U^2 E_k \, , \label{velocity_t} \\
        m_k \, \frac{\rmd \varphi}{\rmd \lambda} &=& \frac{l_k}{U^2 r^2} \, . \label{velocity_varphi} 
    \end{eqnarray}
    
    Furthermore, the canonical momenta satisfy the mass-shell condition:
    
    \begin{equation}\label{mass_shell}
        g^{\mu\nu} \, p^{(k)}_\mu \, p^{(k)}_\nu = m_k^2 \, , 
    \end{equation}
    
    which, upon inserting \rref{cycvar} and solving for $p^{(k)}_r$, gives
    
    \begin{equation}\label{radial_momenta}
        p^{(k)}_r = -m_k \, U^2 \frac{\rmd r}{\rmd \lambda} = \sigma_r U^2 E_k \sqrt{ \Big( 1 - \frac{l_k^2}{E_k^2} \frac{1}{U^4 r^2} - \frac{m_k^2}{E_k^2} \frac{1}{U^2} \Big) } \, ,
    \end{equation}
    
    with $\sigma_r = \pm 1$ distinguishing between the two possible branches of the radial motion: ingoing ($\sigma_r = +1$) and outgoing ($\sigma_r = -1$) trajectories.
    
    At this stage, it is important to establish the methodological approach adopted in this paper. Although the exact trajectory of a massive particle is timelike, neutrinos are ultrarelativistic, and their trajectories can, to leading order, be approximated by the corresponding null geodesic. This prescription is often referred to as the \textit{null approach}. This approximation has proved to be convenient for the calculation of the phase \rref{covariant_phase_wkb} in curved spacetimes, since it allows one to evaluate the propagation path independently of the neutrino mass, while the leading mass dependence is still retained in the phase calculation through the mass-shell condition \rref{mass_shell} \cite{Cardall:1996cd,Fornengo:1996ef}. Consequently, the phase shift between two different mass eigenstates is expanded in powers of the mass, with the leading order being proportional to the mass-squared difference, as in the standard treatment of neutrino oscillations in flat spacetime.
    
    It is worth mentioning here that the use of the null approach is related to a well-known ``factor of two'' issue in the calculation of the neutrino oscillation phase in comparison to a direct evaluation along the corresponding timelike geodesic. In other words, when comparing the phase obtained using the null approach with that obtained by direct integration along the corresponding timelike geodesic, the latter can yield a result that is twice the former at leading order in the $m_k^{\, 2}$ expansion \cite{Zhang:2003pn}. This difference reflects the different prescriptions used to evaluate the phase and the mass dependence of the propagating trajectory and the quantities held fixed in the expansion. Furthermore, since the null approach combines null trajectories with mass-dependent momenta, in the following calculations we will take care to consistently define each physical quantity as it appears in the calculation of the oscillation phase. Hence, we adopt the null approach and retain the leading non-vanishing terms in $m_k^{\, 2}/E_k^2$ in the calculations that follow.
    
    In the null approach, the phase \rref{covariant_phase_wkb} is then \cite{Cardall:1996cd,Fornengo:1996ef}
    
    \begin{equation}
        \Phi^{(k)} = \int_{r_S}^{r_D} \left[ E_k \, \left. \frac{\rmd t}{\rmd r} \right|_{0} - l_k \, \left. \frac{\rmd \varphi}{\rmd r} \right|_{0} + p_r^{(k)} \right] \rmd r \, ,
    \end{equation}
    
    where the quantities: 
    
    \begin{eqnarray}
        \left. \frac{\rmd t}{\rmd r} \right|_{0} &=& - \frac{U^2}{\scriptr_0} \, , \label{conserved_t} \\
        \left. \frac{\rmd \varphi}{\rmd r} \right|_{0} &=& -\frac{b}{U^2 r^2 \scriptr_0} \, , \label{conserved_varphi} 
    \end{eqnarray}
    
    are evaluated along the corresponding null geodesic. Here, we have already introduced the impact parameter along the null trajectory, $b = l_0 / E_0$, and introduced the shortcut: 
    
    \begin{equation}
        \scriptr_0 = \sigma_r \, \sqrt{1-\frac{b^2}{U^4 r^2}} \, .
    \end{equation}
    
    Inserting these into the phase and rearranging the expression, we find: 
    
    \begin{equation}
        \Phi^{(k)} = - \int_{r_S}^{r_D} \rmd r \, \frac{U^2 E_k}{\scriptr_0} \left[ 1 - \frac{l_k}{E_k} \frac{b}{U^4 r^2} - \sigma_r \scriptr_0 \, \sqrt{1 - \frac{l_k^{\, 2}}{E_k^{\, 2}} \frac{1}{U^4 r^2} - \frac{m_k^{\, 2}}{E_k^{\, 2}} \frac{1}{U^2}}  \right] \, .
    \end{equation}
    
    The dihole MP solution is asymptotically flat, and we associate the massive trajectory and the reference null geodesic with the same asymptotic impact parameter
    
    \begin{equation}
        \frac{l_k}{E_k} \approx b \, v_\infty = b \left( 1 - \frac{m_k^{\,2}}{2E_k^{\,2}} \right) \, .
    \end{equation}
    
    Inserting it into the phase leads to a beautiful simplification: the zeroth order in the mass vanishes, and the phase becomes: 
    
    \begin{equation}
        \Phi^{(k)} = - \frac{m_k^{\,2}}{2E_k} \int_{r_S}^{r_D} \frac{1}{\scriptr_0} \rmd r \, .
    \end{equation}
    
    Last but not least, we here adopt the equal energy prescription, according to which the energy in a given trajectory is taken at leading order to coincide with that of the corresponding null geodesic, that is: $E_k \approx E_0 \, + \, \mathcal{O}(m_k^2/E_0)$. Since the phase is already of leading order $m_k^{\, 2}$, any corrections to the energy $E_k$ would exclusively contribute to higher-order terms in the phase. This is also consistent with the flat-spacetime derivation presented above. 
    
    Now, recovering the path label, the phase for a neutrino traveling in the MP geometry is:
    
    \begin{equation}\label{MP_phase}
        \Phi^{(k,p)} = -\sigma_r \, \frac{m_k^{\, 2}}{2E_0} \, \int_{r_S}^{r_D} \frac{1}{\sqrt{1-\frac{b_p^{\, 2}}{U^4 r^2}}} \rmd r \, ,
    \end{equation}
    
    on each radial branch. From here the advantage of the null approximation is clear -- at leading order, expression \rref{MP_phase} exhibits a direct factorization of the mass and path dependences: \(k\) labels the neutrino mass eigenstate, whereas \(p\) labels the respective null trajectory and thereby determines the corresponding value of $b_p$ and, consequently, the value of the integral.    

    \subsection{Purely radial motion}
    
    A particular case directly follows from \rref{MP_phase} -- for $b_p =0$, that is, purely radial motion, the integration can be straightforwardly performed, yielding 
    
    \begin{equation}\label{phase_radial_motion}
        \Phi^{(k)} = \frac{m_k^{\, 2}}{2 E_0} \lvert r_D - r_S \rvert\, ,
    \end{equation}
    
    which resembles the form obtained in a flat spacetime \rref{flat_phase}, as well as the radial case in other static black hole geometries such as Schwarzschild \cite{Cardall:1996cd,Fornengo:1996ef}, and Reissner-Nordstr\"om \cite{Alexandre:2025qip}. Eq. \rref{phase_radial_motion} requires some attention. At first glance, this may appear to be independent of the spacetime geometry, since it contains no explicit dependence on the metric function $U$; therefore, its functional form is unchanged under the limit $M \to 0$. However, one needs to remember that the coordinate representation as well as the definition of the Killing energy are strictly linked to the spacetime. The energy $E_0$ appearing in \rref{MP_phase} is a conserved quantity along a null geodesic in the MP background from the neutrino's production at the source until its detection. At the detector, the locally measured energy is:
    
    \begin{equation}
        E_{(loc)} = p^{(k,p)}_\mu \, u_D^{\mu} \, , 
    \end{equation}
    
    where $u_D^{\mu}$ is the detector's four-velocity. Note that, from a measurement perspective, one may, for instance, associate the detector with a tetrad reference frame whose temporal basis vector is $ e_{\hat{t}}^{\mu} \equiv u_{D}^\mu$. In this frame, the locally measured energy corresponds to the temporal component of the neutrino four-momentum $p_{\hat{t}} = E_{(loc)} = p^{(k,p)}_\mu e_{\hat{t}}^{\, \mu} $. 
    For the case of a static observer in the MP background, $u_D^{\mu} = \Big( U(r_D),0,0,0 \Big)$, and therefore:
    
    \begin{equation}\label{local_energy}
        E_{(loc)} \simeq U(r_D) \, E_0 \, .
    \end{equation}
    
    Thus, while $E_0$ is the conserved energy along the corresponding null geodesic, the locally measured energy at the detector depends on the redshift factor $U(r_D)$.
    Inserting it directly into the phase, we have:
    
    \begin{equation}\label{phase_MP_locally_energy}
        \Phi^{(k)} = \frac{m_k^{\, 2} U(r_D)}{2E_{(loc)}} \lvert r_D - r_S \rvert \, .
    \end{equation}
    
    Furthermore, note that the \textit{points} $r_S$ and $r_D$ are merely coordinate locations of the source and detector, respectively; and thus $r_D-r_S$ should not be understood as the physical distance between the source and the detector. For a purely radial curve on a constant-time hypersurface, the corresponding proper distance, assuming $r_D>r_S$ is 
    
    \begin{equation}\label{proper_distance}
        L_p = \int_{r_S}^{r_D} \sqrt{-g_{rr}} \, \rmd r = r_D - r_S + 2M \left( \arcsinh{\left( \frac{r_D}{a} \right) } - \arcsinh{\left(\frac{r_S}{a}\right)} \right)  \, .
    \end{equation}
    
    It is important to stress here that $L_p$ represents the proper spatial distance between the source and detector locations on a constant-time hypersurface and should not be understood as the distance traveled by the neutrino along the null geodesic. Particularly, the quantity $\lvert r_D - r_S\rvert$ is a coordinate separation that relates to the proper distance via \rref{proper_distance}. Consequently, although the phase for the purely radial case assumes an analogous form as in the flat-spacetime counterpart when expressed in terms of trajectory energy ($E_0$), the influence of the MP geometry remains encoded in the relation between the coordinates and proper distance separation and, from a more observational perspective, in the relation between the conserved and locally measured energies.

    \section{Gravitational lensing of neutrinos}\label{sec:lensing}
    
    We now turn our attention to the general non-radial case $b_p \neq 0$. In contrast to the purely radial motion, the spacetime geometry plays a direct role in the null-geodesic description and hence in the phase oscillation. Of particular interest is the gravitational-lensing configuration where the neutrino trajectory bends due to the presence of the two black holes. In general, the null-geodesic system on the equatorial plane is described by equations \rref{conserved_t}, \rref{conserved_varphi} in addition to the radial equation:
    
    \begin{equation}\label{radial_velocity_squared}
        \frac{1}{E_0^{2}} \left( \frac{\rmd r}{\rmd \lambda} \right)^2 = 1 - \frac{b_p^{\, 2}}{U^4 r^2} = 1 - b_p^{\, 2} \, V_{eff}(r) \, \geq 0 ,
    \end{equation}
    
    where the condition on the right-hand side determines the physically relevant orbits. Furthermore, we introduce the effective potential on the equatorial plane as:
    
    \begin{equation}\label{effective potential}
        V_{eff}(r) = \frac{1}{U^4 r^2} \, .
    \end{equation}
    
    This effective-potential approach is particularly useful for classifying geodesic motion in curved spacetimes, as it keeps a clear analogy with classical mechanics. Thus, the qualitative behavior of the null geodesics can be obtained from the structure of $V_{eff}$. For the gravitational-lensing configurations considered here, an ultrarelativistic neutrino emitted by a distant source approaches the gravitational center, reaches the point of closest approach $r_0^{p}$, and then propagates towards the detector. At the turning point, the corresponding null geodesic satisfies  
    
    \begin{equation}\label{impact_parameter_closest}
        b_p^{\, 2} = \frac{1}{V_{eff}(r_{0})} = U(r_{0})^{4} r_{0}^2 \, .
    \end{equation}
    
    Furthermore, the structure of the null geodesics in the equatorial plane for the dihole MP solution is rich, and thus the same can be said for gravitational lensing. Particularly, there may exist circular null geodesics, known as \textit{light rings} (LRs), located at radii $r_{LR}$, satisfying 
    
    \begin{equation}
        \frac{\rmd}{\rmd r}V_{eff}(r_{LR}) = 0 \, , \ \ \ \qquad \ \ \ b_{LR}^2 = \frac{1}{V_{eff}(r_{LR})} = r_{LR}^2 \, U(r_{LR})^{4} \, .
    \end{equation}
    
    The existence of LRs in the symmetry plane, while the two BHs are located along the \(z\)-axis, is rather interesting. Depending on the ratio $M/a$, the equatorial plane can admit two, one degenerate, or no LRs at all. In other words, there is a critical value $\eta_{crit}= \sqrt{27/8}$ that separates three different scenarios. For $M/a > \eta_{crit}$, two LRs exist, corresponding to an unstable outer LR and an inner stable LR. As the distance between the two BHs increases, the ratio $M/a$ decreases, reaching the critical value $M/a = \eta_{crit}$, in which both LRs degenerate into a single one located at $r^2 = 5 a^2$. For $M/a < \eta_{crit}$, no LR exists in the equatorial plane \cite{MP_lensing_complete}.  
    
\begin{figure}[h]
        \centering
        \includegraphics[width=0.5\linewidth]{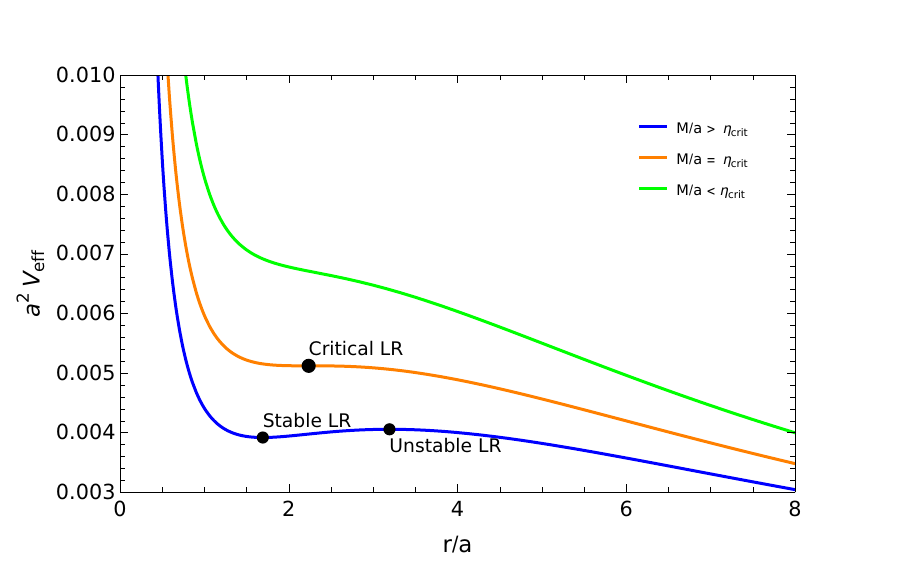}
        \caption{Schematic representation of the effective potential for the equal-mass dihole MP solution in the equatorial plane. The existence of LRs is strongly related to the ratio between the black hole's individual mass, $M$, and the half-separation $a$. For $M/a > \eta_{crit}$ (blue), there are two LRs. For $M/a = \eta_{crit}$ (orange), the two LRs degenerate into a single one. For $M/a < \eta_{crit}$ (green), there are no LRs in the equatorial plane.
        }
        \label{fig:effective_potential}
    \end{figure}

    For isolated BHs, infinitely many relativistic gravitational-lensing images can arise from null trajectories whose turning points approach an unstable LR from outside. Interestingly, for the dihole MP configuration considered here, a null geodesic can penetrate the unstable outer LR and then, due to the subsequent increase of the effective potential in the inner region, reach an inner turning point, where it is turned back and eventually escapes towards the detector, leading to a different set of gravitational-lensing images \cite{MP_lensing_complete}. Consequently, the gravitational-lensing phenomenology differs considerably among the three possible configurations. Since, within the null approximation, the neutrino propagation path is approximated by a corresponding null geodesic, these distinct lensing trajectories can give rise to different path-dependent contributions to the neutrino oscillation phase. The behavior of the effective potential for these three cases is presented in \figref{fig:effective_potential}. 
    
    Having settled the general gravitational-lensing dynamics of the dihole MP geometry, we turn now to the relevant trajectories connecting the source and detector. In general, the accumulated change of the azimuthal coordinate along a given path \(\gamma_p\) can be taken as
    
    \begin{equation}
        \Delta \varphi = \int_{\gamma_p} \rmd \varphi = \int_{\gamma_p} \left. \frac{\rmd \varphi}{\rmd r} \right|_0 \rmd r = \varphi_D - \varphi_S + 2n\pi \, ,
    \end{equation}
    
    where $n \in \mathbb{Z} $ accounts for possible windings around the gravitational center. Therefore, the same source and detector positions can be connected by different trajectories characterized by different accumulated angular displacements. In the following, we restrict our analysis to the primary lensing trajectories, that is, the null geodesics connecting the source and detector without additional windings around the gravitational center.  
    
    Particularly, we consider the case of a neutrino emitted by a distant source located at (\(r_S,\varphi_S\)) which propagates inwards until it reaches the point of closest approach $r_0^{p}$, and subsequently propagates outwards until reaching the detector, located at (\(r_D,\varphi_D\)). Thus, splitting each trajectory into ingoing and outgoing branches, the angular displacement is 
    
    \begin{equation}\label{angular_displacement}
        \lvert \Delta\varphi \rvert = \int_{r_0^{p}}^{r_S} \frac{\lvert b_p\rvert} {U^2r^2 \lvert\scriptr_0\rvert} \, \rmd r + \int_{r_0^{p}}^{r_D} \frac{\lvert b_p\rvert}{U^2r^2 \lvert\scriptr_0\rvert} \, \rmd r \, .
    \end{equation}
    
\begin{figure}[h!]
        \centering
        \includegraphics[width=0.75\linewidth]{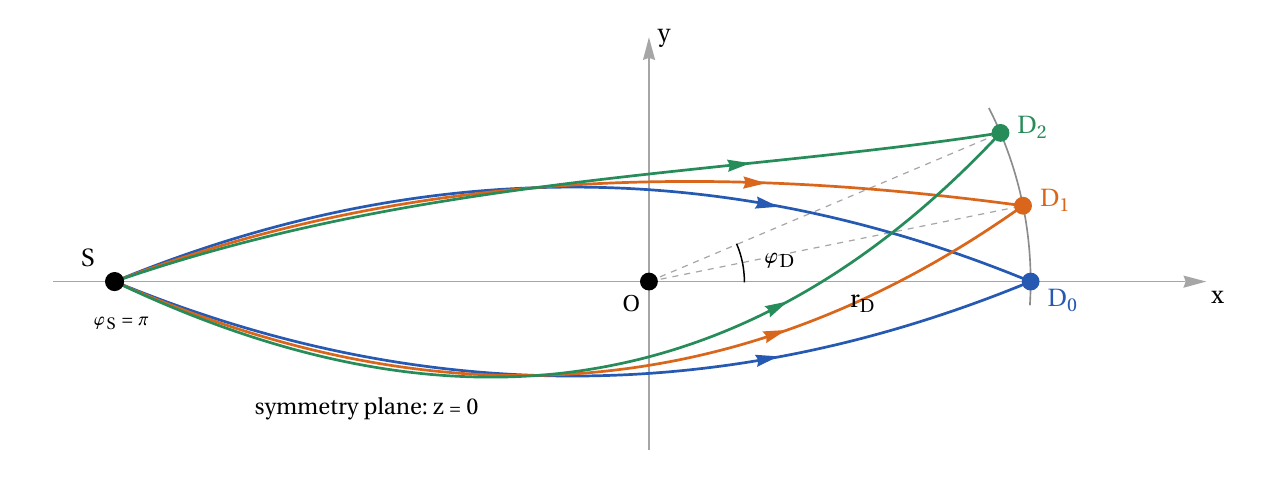}
        \caption{Schematic representation of the gravitational lensing in the symmetry plane \(z=0\). For each detector angular position, denoted by \(D_0\), \(D_1\), and \(D_2\), two paths connect the fixed source to the detector at the same radius \(r_D\). 
        }
        \label{fig:2path_schematic}
    \end{figure}

    Here, \(\lvert \Delta\varphi \rvert\) represents the magnitude of the angular displacement accumulated along the trajectory, whereas the sign of \(b_p\) determines the orientation around the gravitational center. The impact parameter relates to the point of closest approach through \rref{impact_parameter_closest}. Therefore, for fixed source and detector positions, the angular displacement equation \rref{angular_displacement} can be used to obtain the point of closest approach $r_0^{p}$ associated with a given path. For the gravitational-lensing settings considered below, we fix the source at $\varphi_S = \pi$, while the detector is on the other side of the gravitational lens; we set $\varphi_D \geq 0$. Two schematic paths connecting the source to each of three detector angular positions are illustrated in \figref{fig:2path_schematic}. 
    
    The numerical setup applied here is the following. For each considered position of the detector, we solve the angular displacement equation \rref{angular_displacement}, obtaining the respective turning point \(r_0^{p}\) and impact parameter \(b_p\) for each path. With all trajectories determined over the considered range of the detector's angular position, the phase along them can be evaluated for every \(k\)-th mass eigenstate. For trajectories containing a single turning point, \rref{MP_phase} becomes 
    
    \begin{equation}
        \Phi^{(k,p)} = \frac{m_k^{\, 2}}{2E_0} \left[ \int_{r_0^{p}}^{r_S} \frac{1}{\lvert \scriptr_0 \rvert} \rmd r + \int_{r_0^{p}}^{r_D} \frac{1}{\lvert \scriptr_0 \rvert} \rmd r \right] \, .
    \end{equation}
    
    We first consider the case in which the spacetime has two LRs on the symmetry plane, that is, \(M/a > \eta_{crit}\), and we restrict ourselves to the two primary trajectories whose turning points lie outside the outer LR. We consider the two impact parameters to have opposite signs according to the orientation of each trajectory around the gravitational center. Consequently, considering an electron neutrino produced at the source with the flavor state
    
    \begin{equation}
        \ket{\nu_e} = \cos\theta \ket{\nu_1} + \sin\theta \ket{\nu_2} \, ,  
    \end{equation}
    
    the probability of obtaining a muon neutrino after propagating along the two primary gravitational-lensing paths takes the compact form
    
    \begin{equation}\label{two_flavors_probability}
        P_{e \rightarrow \mu} = \frac{\lvert\mathcal{N}\rvert^2}{4} \sin^2 2\theta \, \lvert \mathcal{P}_2 - \mathcal{P}_1 \rvert^2 \, , 
        \qquad
        \lvert \mathcal{N} \rvert^2 = \frac{1}{\cos^2 \theta \lvert \mathcal{P}_1 \rvert^2 + \sin^2\theta \lvert \mathcal{P}_2 \rvert^2 } \, .
    \end{equation}
    
    The path-propagation factors, \( \mathcal{P}_1 \) and \( \mathcal{P}_2 \), are given by \rref{Path_shortcut}. To demonstrate how different interference contributions affect the resulting transition probability, namely those between different mass eigenstates propagating along the same trajectory, the same mass eigenstate propagating along different trajectories, and different mass eigenstates propagating along different trajectories, we explicitly write the quantities entering the two-path, two-flavor transition probability as
    
    \begin{eqnarray}\label{two_path_two_flavor_probability_explicitly}
        \lvert \mathcal{P}_2 - \mathcal{P}_1 \rvert^2 &=& 4 + 2\cos\left( \Phi^{(1,1)}-\Phi^{(1,2)} \right) + 2\cos\left( \Phi^{(2,1)} -\Phi^{(2,2)} \right) - 2\cos\left( \Phi^{(2,1)}-\Phi^{(1,1)} \right) \\ 
        && -\, 2\cos\left( \Phi^{(2,1)}-\Phi^{(1,2)} \right) - 2\cos\left( \Phi^{(2,2)}-\Phi^{(1,1)} \right) - 2\cos\left( \Phi^{(2,2)}-\Phi^{(1,2)} \right) \, , \nonumber \\
        \lvert \mathcal{N}\rvert^2 &=& \frac{1}{2\cos^2\theta \left[ 1+ \cos\left( \Phi^{(1,1)} - \Phi^{(1,2)} \right) \right] + 2\sin^2\theta \left[ 1+ \cos\left( \Phi^{(2,1)} - \Phi^{(2,2)} \right) \right] } \, \label{two_path_normalization} ,
    \end{eqnarray}
    
    Hence, the gravitational-lensing trajectories play a fundamental role in both the flavor-transition probability and the normalization of the propagated state. The differences \( \Phi^{(1,p)} - \Phi^{(2,p)} \) compare different mass eigenstates propagating along the same trajectory and are proportional to the mass-squared difference, thus recovering the flat limit factor of \(\Delta m^2/E\).    
    
    In contrast to the standard two-flavor oscillation probability, the interference between different paths introduces a direct dependence on the individual neutrino mass eigenvalues even in the null approach. Consequently, for non-maximal mixing, the transition probability can also become sensitive to the sign of \(\Delta m^2 = m_2^{\,2} - m_1^{\,2} \) \cite{Swami:2020qdi}. In fact, the interchange of the two mass eigenstates, accompanied by the transformation \(\theta \to \pi/2-\theta \), leaves the transition probability invariant. Hence, for maximal mixing, \(\theta = \pi/4\), the latter transformation leaves the mixing angle unchanged and thus, the distinction associated with the sign of \(\Delta m^2\) disappears. Note also that, within the assumed formalism, the normalization depends, for each neutrino mass eigenstate, on the interference arising due to propagation along different trajectories. Consequently, variation in the detector position modifies the trajectories and their corresponding phases, therefore affecting the observed probability. 
    
    The configurations considered below, described by the dihole MP solution, are not intended to represent any particular astrophysical binary system. Nevertheless, we choose to set the parameter values on astrophysical scales. For simplicity, we restrict the numerical analyses to two flavors, although the construction extends to three flavors as well. We follow the standard nomenclature and distinguish between two mass orderings depending on the sign of \( \Delta m^{2} = m_2^{\, 2} - m_1^{\, 2} \). The lightest neutrino mass is set to zero, whereas the remaining neutrino mass squared is taken as the adopted \( \lvert \Delta m^2 \rvert \). Thus, for \( \Delta m^2 > 0 \), we take \(m_1 = 0\) and \( m_2^{\, 2} = \lvert \Delta m^2 \rvert \), and the mass assignments are interchanged for \(\Delta m ^2 <0\). The same mass assignments have been used in two-flavor models of gravitationally lensed neutrino oscillations  \cite{Swami:2020qdi,Chakrabarty:2023kld,Alloqulov:2024sns,Wang:2025nyl,AraujoFilho:2025rzh}. 
    
    So far in this manuscript, we have adopted natural units, that is \( G=c=\hbar=1 \). For the numerical analysis, we choose to present the results in units commonly used in high-energy physics. This means that, for the numerical simulation, our input mass is taken as the geometrized length \( M \to GM/c^2 \). Hence, \(M\), \(a\) and all radial coordinates are given in km, while the energy and neutrino mass are expressed in \(eV\). Accordingly, the neutrino phase \rref{MP_phase} factor needs to be rescaled properly as \( m_k^{\, 2}/E_0 \to m_k^{\, 2}/(E_0 \hbar c) \) so that the phase remains dimensionless.   
    
\begin{figure}[h]
        \centering
        \includegraphics[width=0.45\linewidth]{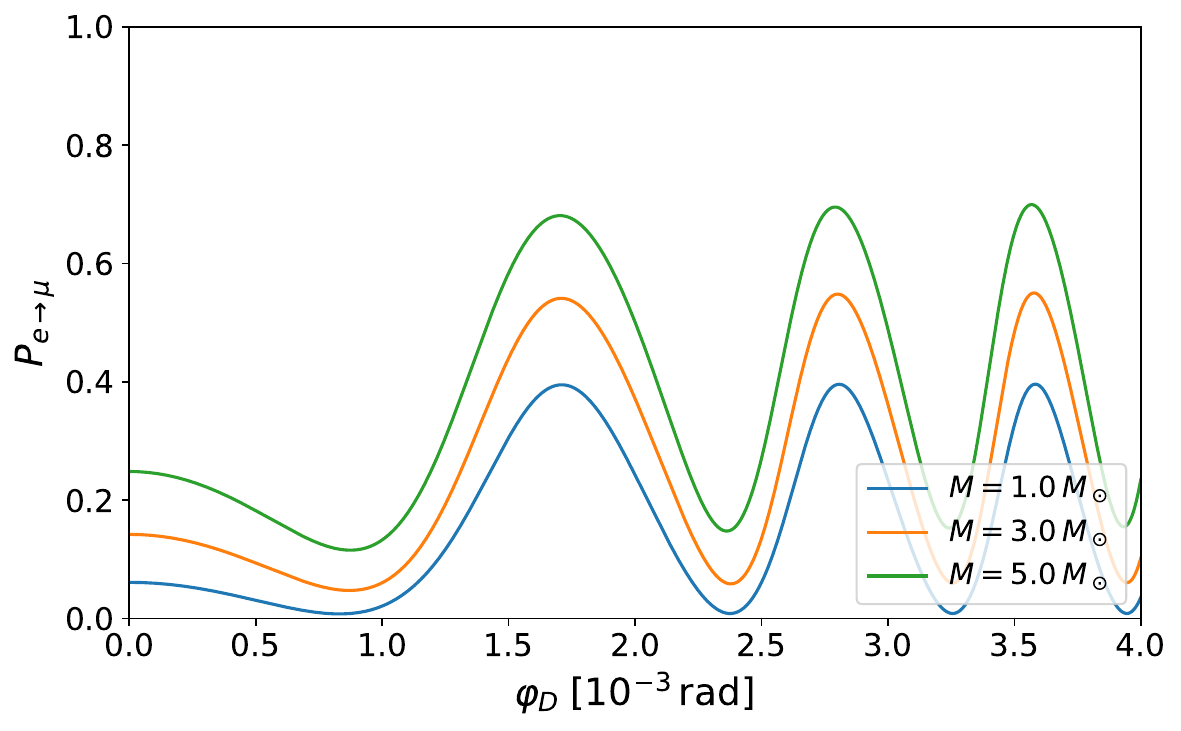}
        \includegraphics[width=0.45\linewidth]{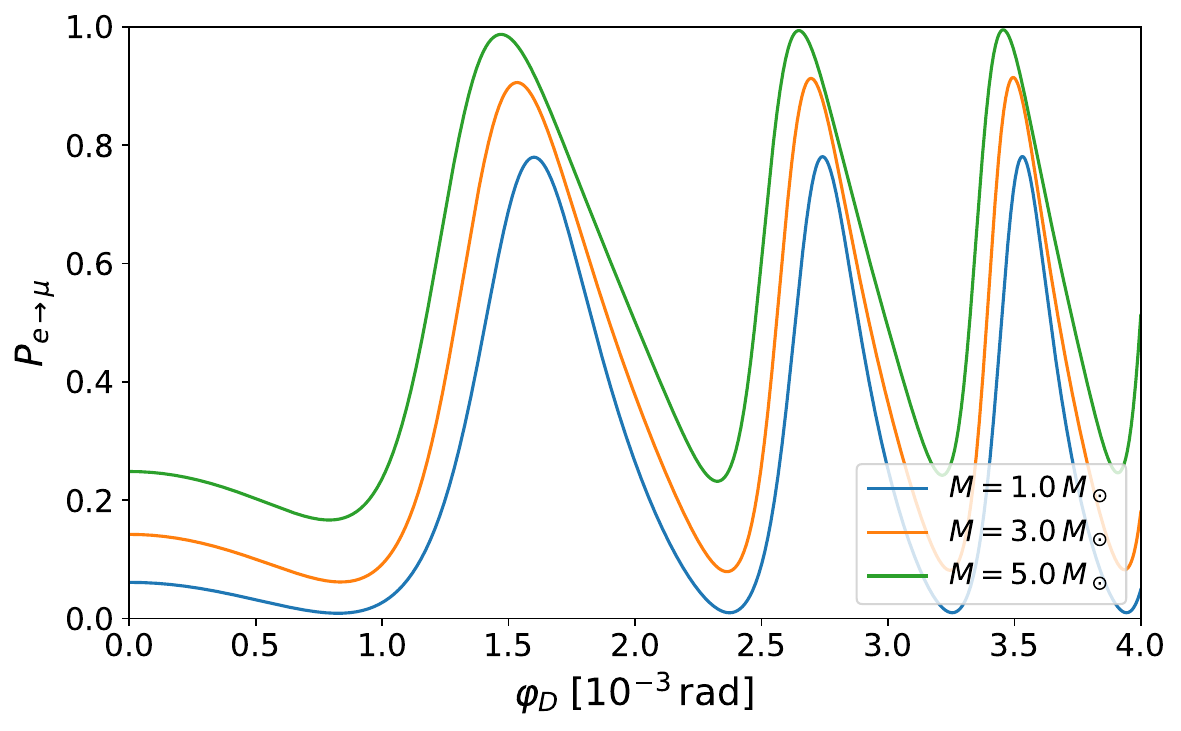}
        \caption{Two-flavor transition probability along two primary paths in the weak-lensing regime of the MP dihole geometry. The source is fixed at $(r_S,\varphi_S)=(10^5r_D,\pi)$, while the detector position varies over \(0 \leq \varphi_D \leq 4 \times 10^{-3}\) rad at fixed $r_D=10^8\,\mathrm{km}$. We set $E_0=10^8\,\mathrm{eV}$, $|\Delta m^2|=10^{-3}\,\mathrm{eV}^2$, $\theta=33.65^\circ$, and the half-separation $a=M_\odot/3$. The curves correspond to black-hole masses $M=\{1,3,5\}M_\odot$. The left panel has $\Delta m^2>0$ ($m_1=0$), and the right panel has $\Delta m^2<0$ ($m_2=0$).}
        \label{fig_weak_lensing}
    \end{figure}

    In our numerical setup, we consider the complete integration of the curves connecting the source and detector; still, the first case we study is actually the weak lensing. In \figref{fig_weak_lensing} we consider the detector moving with constant $r_D = 10^{8} $km, while its angular position varies over $ 0 \leq \varphi_D \leq \, 4 \times 10^{-3}$. The source is fixed at $r_S = 10^{5} \, r_D$, $\varphi_S=\pi$. The neutrino moves along a trajectory with conserved energy $E_0 =10^8$ eV. We keep the halfway distance between the black holes fixed at $a = M_{\odot}/3$ and consider three different masses $M = \{1,3,5 \} M_{\odot}$, where $M_\odot$ denotes the geometrized solar mass. All of these scenarios then have two LRs on the equatorial plane. We have numerically verified that all plotted trajectories satisfy: \( M, a \ll r_0^{p} \approx \lvert b_p \rvert \ll r_D < r_S \), thus characterizing the weak-lensing regime. Hence, for all trajectories in \figref{fig_weak_lensing}, the point of closest approach lies outside the outer LR. At \( \varphi_D=0 \), the two primary paths have equal phase integrals, and thus the transition probability agrees between the two mass orderings for each considered value of \(M\). As the detector moves away from the alignment, the unequal paths produce distinct phase contributions. The left panel corresponds to \(\Delta m^2 >0\) while the right panel corresponds to \(\Delta m^2<0\). For the same geometry and \(\lvert \Delta m^2 \rvert\), interchanging the mass ordering leaves the numerator of \rref{two_flavors_probability} unchanged, but exchanges the two path factors in its normalization. With the chosen nonmaximal mixing angle \( \theta=33.65^\circ < 45^\circ \), the \(\Delta m^2 < 0\) probability is greater than or equal to the \(\Delta m^2 >0\) counterpart at each detector's position. Furthermore, increasing \(M\), within the considered range of \(\varphi_D\), the curves relative to larger black hole mass generally access larger transition probabilities. We also noted that a small variation in the probability peak position appears at smaller values of \(\varphi_D\) as the values of \(M\) increase; this effect is more clearly observed in the \(\Delta m^2 <0\) case shown in the right panel.     
    
\begin{figure}[h]
        \centering
        \includegraphics[width=0.45\linewidth]{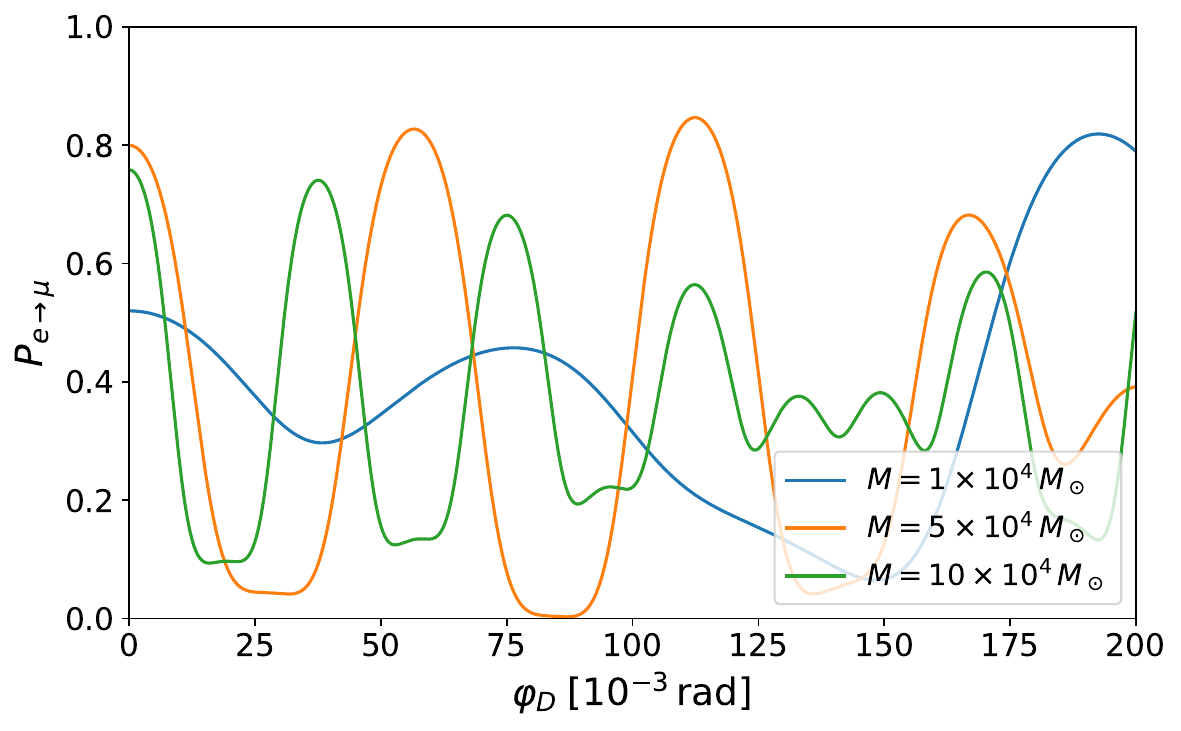}
        \includegraphics[width=0.45\linewidth]{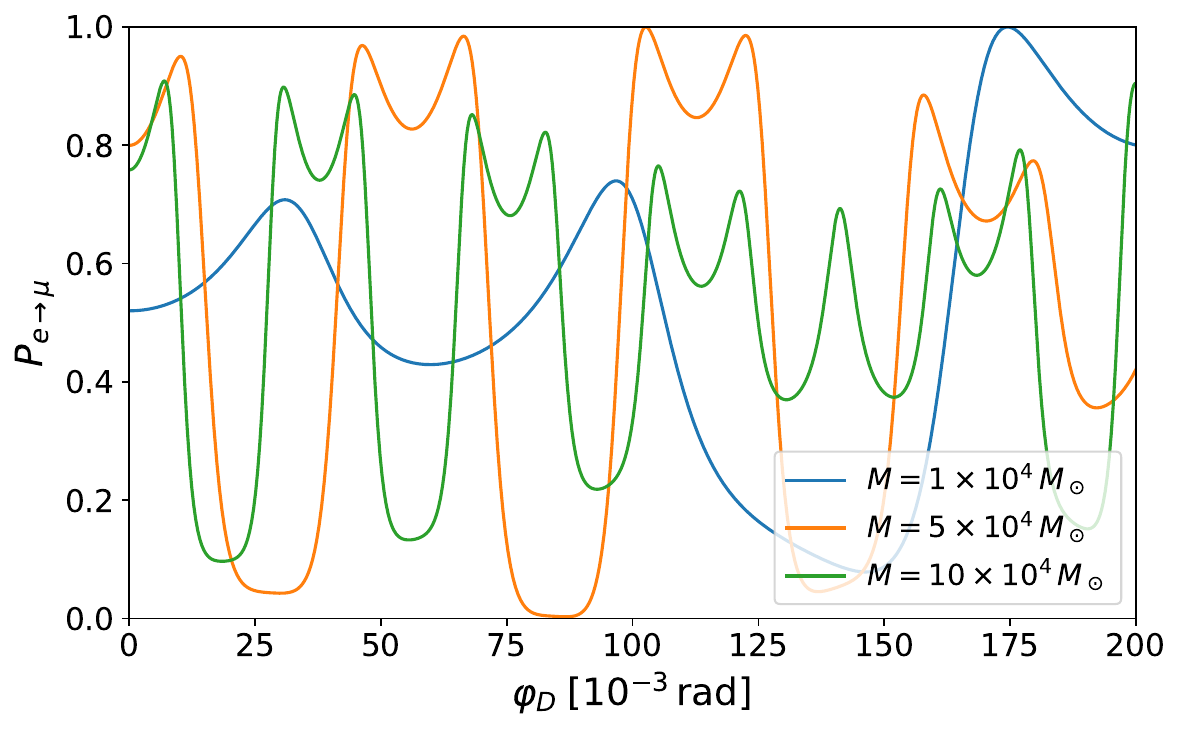}
        \caption{Two-flavor transition probability along two zero-winding paths in stronger-lensing configurations of the MP dihole geometry. The source is fixed at $(r_S,\varphi_S)=(10^3r_D,\pi)$, while the detector position varies over $0\leq\varphi_D\leq0.2$ rad at fixed $r_D=10^6\mathrm{km}$. We set $E_0=10^{10}\mathrm{eV}$, $|\Delta m^2|=10^{-3}\mathrm{eV}^2$, $\theta=33.65^\circ$, and the half-separation $a=M_\odot/3$. The curves correspond to black-hole masses $M=\{1,5,10\}\times10^4M_\odot$. The left panel has $\Delta m^2>0$ ($m_1=0$), and the right panel has $\Delta m^2<0$ ($m_2=0$). Both paths turn outside the outer LR.}
        \label{fig:strong_lens_2LR}
    \end{figure}

    As we increase the value of each black hole mass, and therefore the gravitational-lensing effect, the transition probability pattern changes significantly. Since the neutrino phases scale with \(E_0^{-1}\), the adopted energy strongly affects the number of oscillations in a given angular interval. In \figref{fig:strong_lens_2LR} we show the probability for configurations with a stronger lensing effect. The source is fixed at \( \{ r_S, \varphi_S \} = \{ 10^3\,r_D, \pi \} \), while the detector remains at \(r_D=10^6 \mathrm{km}\), with its angular position varying over $0 \leq \varphi_D \leq 0.2 $ rad, a considerably wider interval than in the weak lensing case. The half-separation between the two BHs is kept fixed at \(a=M_\odot/3\), and we adopt \(E_0=10^{10}\mathrm{eV} \). As expected, at \(\varphi_D=0\), the probabilities for the two mass orderings agree for each value of \(M\). As the detector moves away from alignment, the two paths acquire different phase integrals and hence differences in the probabilities for the two orderings become apparent. All three considered configurations possess two LRs on the equatorial plane, and we verified numerically that the point of closest approach for each path remains outside the outer LR. Within the considered angular interval of \( \varphi_D \), increasing \(M\) produces more closely spaced oscillations and changes the positions of the peaks. More noticeably than in the weak-lensing case, successive maxima and minima occur at different probability values, reflecting changes in the mean phase, the relative phase between the two paths, and the normalization as the angular position of the detector varies. As in the weak-lensing case, the probability for \(\Delta m^2 < 0\) (right panel) is greater than or equal to that for the \(\Delta m^2 >0 \) (left panel) at each detector position for the same \(M\). Although the gravitational lensing is stronger in these configurations, our numerical analysis is restricted to the two zero-winding paths with turning points outside the outer LR.
    
\begin{figure}[h]
        \centering
        \includegraphics[width=0.45\linewidth]{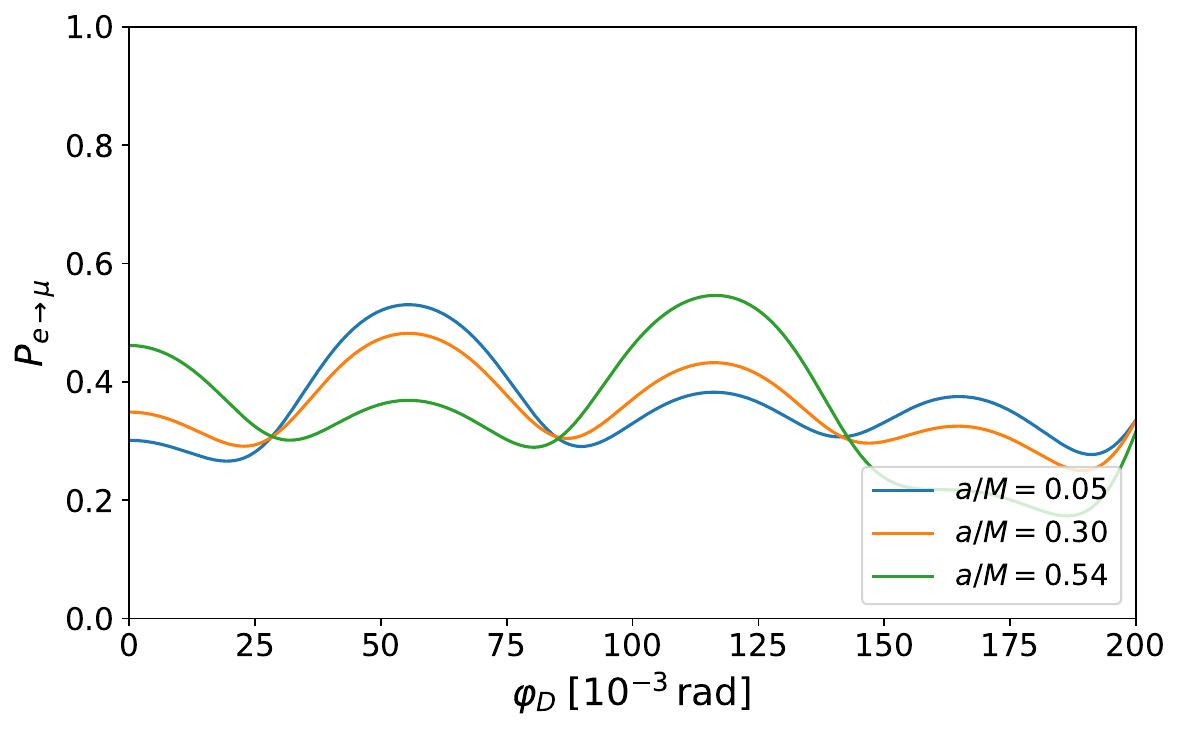}
        \includegraphics[width=0.45\linewidth]{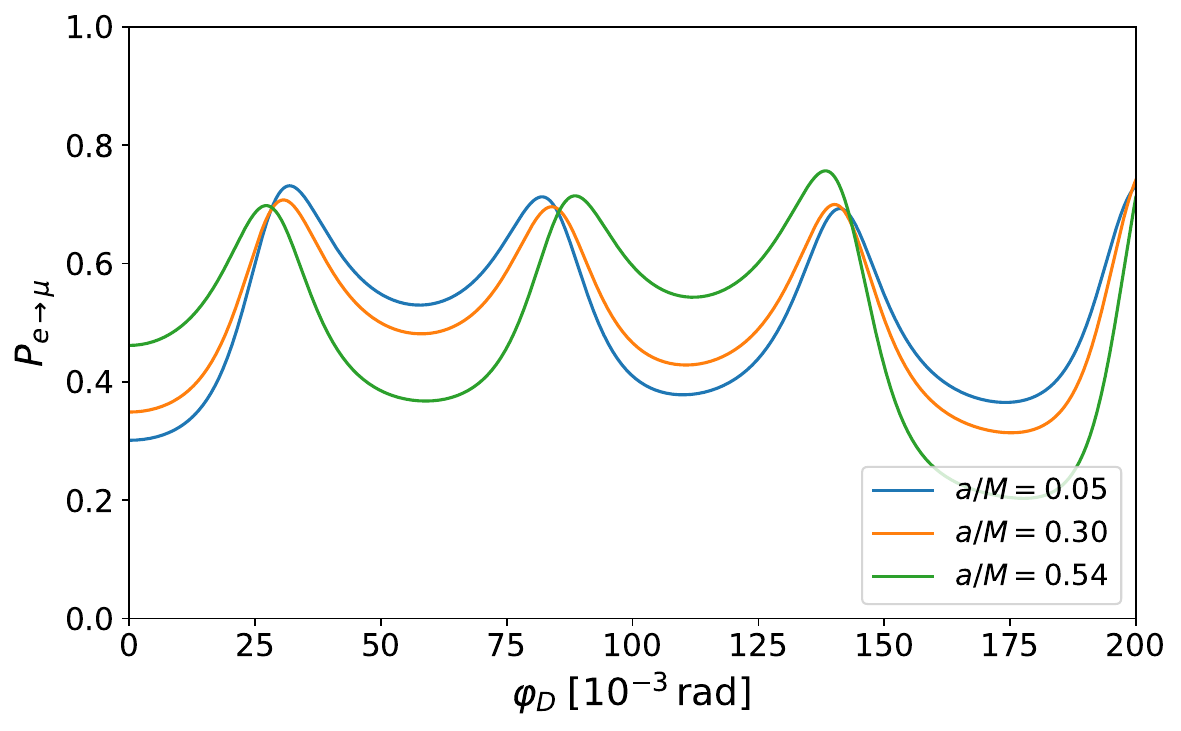}
        \caption{Two-flavor transition probability along two primary paths in the stronger-lensing configurations of the MP dihole geometry for different half-separations. The source is fixed at $(r_S,\varphi_S)=(10^3r_D,\pi)$, while the detector position varies over $0\leq\varphi_D\leq0.2$ rad at fixed $r_D=10^6\,\mathrm{km}$. We set $M=10^5M_\odot$, $E_0=3\times10^{10}\,\mathrm{eV}$, $|\Delta m^2|=10^{-3}\,\mathrm{eV}^2$, and $\theta=33.65^\circ$. The curves correspond to $a/M=\{0.05,0.30,0.54\}$. The left panel has $\Delta m^2>0$ ($m_1=0$), and the right panel has $\Delta m^2<0$ ($m_2=0$). All plotted trajectories turn outside the outer LR.}
        \label{fig:strong_lens_diff_a}
    \end{figure}

    We now examine the effect of changing the half-separation \(a\) at fixed \(M\). Increasing \(a\) weakens the gravitational effect on the symmetry plane while bringing the configurations closer to the critical value at which the two LRs merge. In the weak-lensing regime, we could not discern a change in the transition probability as the half-separation varied. This is expected since the trajectories turn at a considerable distance from the dihole, suggesting that the dependence on \(a\) may become apparent in stronger gravitational configurations. To understand this dependence, we consider a larger mass, \(M=10^5 M_\odot\), and choose values of \(a\) for which the configurations still retain two LRs on the equatorial plane. The resulting curves in \figref{fig:strong_lens_diff_a} retain a similar oscillatory pattern as \(a\) approaches the critical value, but the heights and positions of the extrema change. These differences do not correspond to a uniform increase or decrease in the probability with changing the separation. Although we consider a stronger gravitational lens, we numerically verified that the turning points remain well outside the outer LR for all the trajectories considered. For the parameters considered here, our results suggest that changing the half-separation modulates the probability along the existing oscillatory pattern without substantially changing the overall sequence of oscillations.
    
    We now turn to the critical configuration, in which the two LRs merge, and the configurations in which they disappear on the equatorial plane -- corresponding to \(M/a = \eta_{crit} = \sqrt{27/8}\) and \(M/a < \sqrt{27/8}\), respectively. Similarly to the cases discussed above, we consider different values of \(a\) at fixed \(M\) and different values of \(M\) at fixed \(a\), keeping the same critical configuration as a reference. It is important to highlight that restricting the turning points to the region outside the outer LR does not necessarily include all primary trajectories connecting the source and detector. Configurations possessing LRs can admit infinitely many connecting rays when additional windings are considered \cite{MP_lensing_complete}. Here, however, we restrict our discussion to the zero-winding trajectories. For the configurations considered with \(M/a \leq \eta_{crit}\), we found numerically that there are three trajectories connecting the source and detector throughout the considered interval. In addition to the two outer trajectories, there is a third one passing near the symmetry axis. For three paths, the normalization and transition probability are
    
    \begin{eqnarray}
        P_{e \to \mu} &=& \frac{\lvert\mathcal{N}\rvert^2}{2} \sin^2 2\theta \Bigg[ 3 - \sum_{p=1}^{3} \cos{\left(\Phi^{(2,p)} - \Phi^{(1,p)} \right) } +
        \sum_{p<q} \Big\{ \cos{ \left( \Phi^{(1,p)} - \Phi^{(1,q)} \right) } \\
        &&  + \cos{ \left( \Phi^{(2,p)} - \Phi^{(2,q)} \right) } - \cos{ \left( \Phi^{(2,p)} - \Phi^{(1,q)} \right) } - \cos{ \left( \Phi^{(2,q)} - \Phi^{(1,p)}  \right) } \Big\}  \Bigg] \, , \nonumber \\
        \lvert\mathcal{N}\rvert^2 &=& \left[ 3 + 2 \sum_{p<q} \left\{ \cos^2\theta \cos{\left( \Phi^{(1,p)} - \Phi^{(1,q)} \right)} + \sin^2\theta \cos{\left( \Phi^{(2,p)} - \Phi^{(2,q)} \right)} \right\}  \right]^{-1} \, .
    \end{eqnarray}
    
    Here, the sum over \(p<q\) runs over the three distinct path pairs. 

        \begin{figure}[h]
        \centering
        \includegraphics[width=0.45\linewidth]{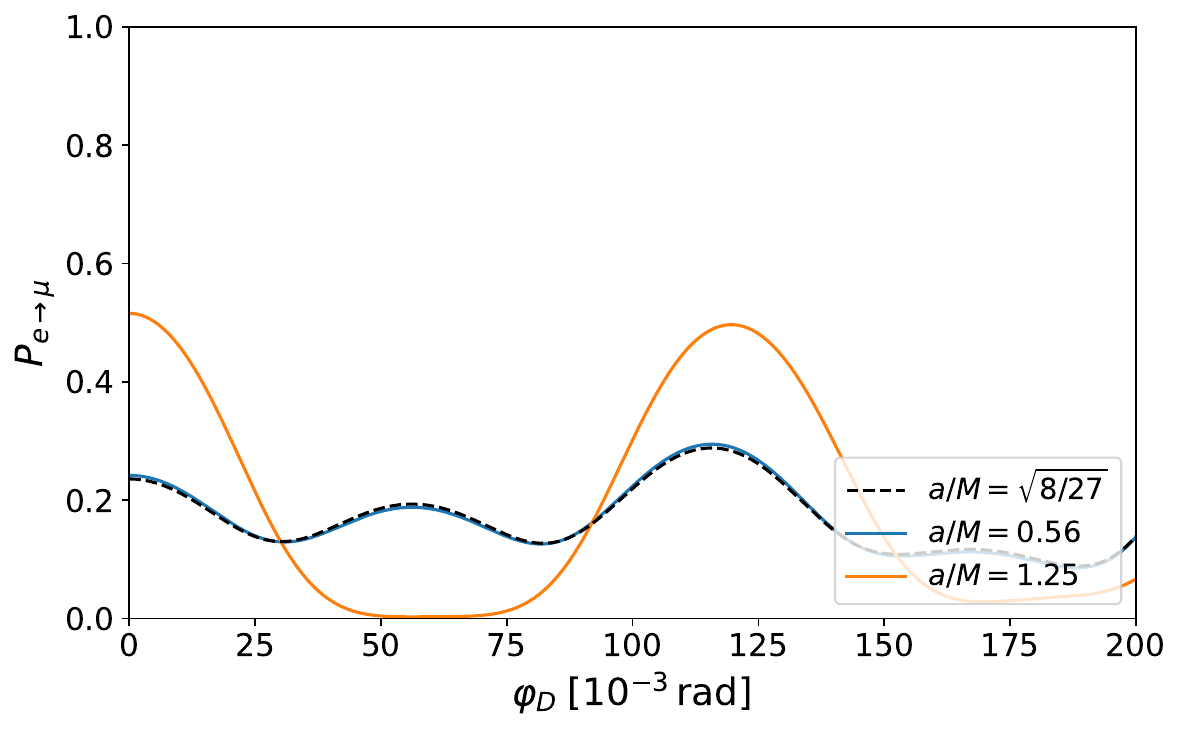}
        \includegraphics[width=0.45\linewidth]{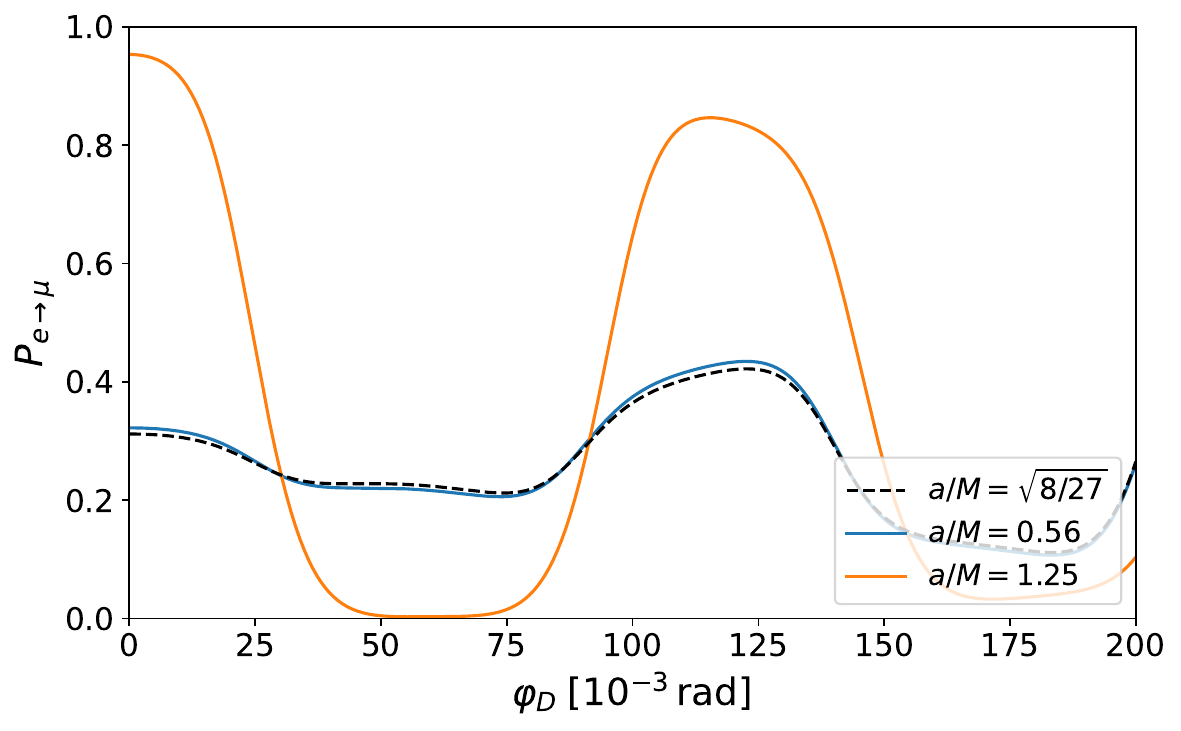}
        \includegraphics[width=0.45\linewidth]{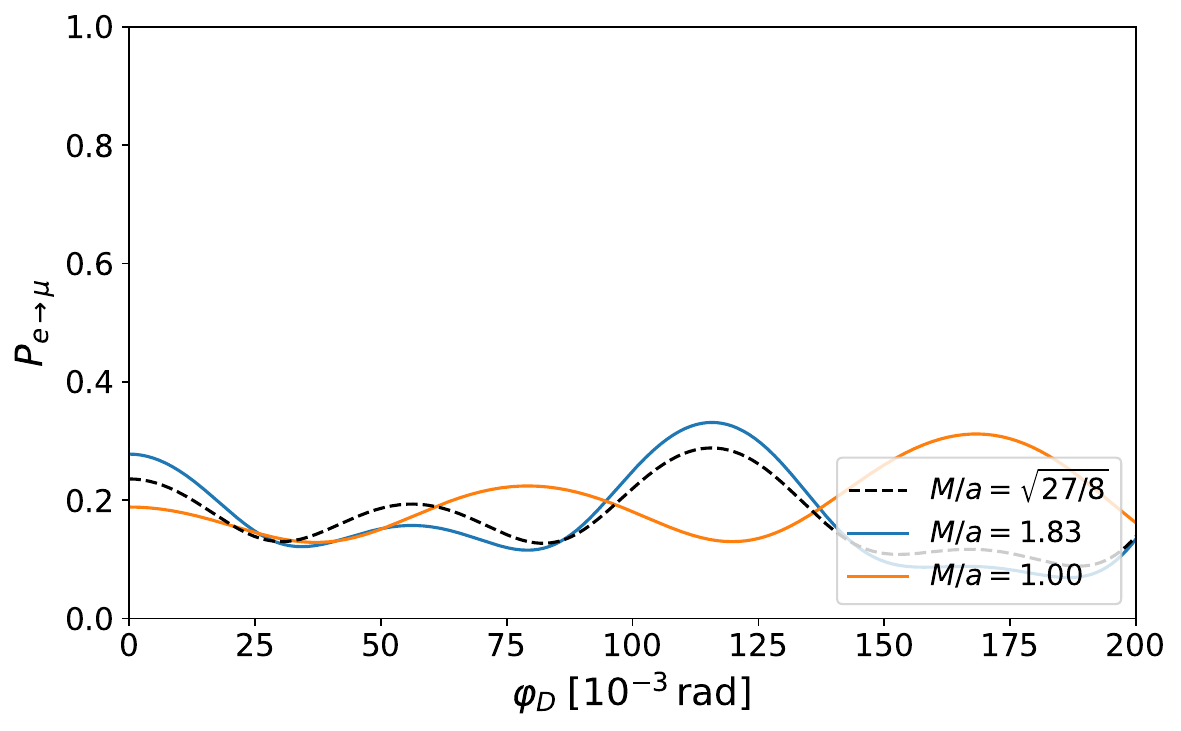}
        \includegraphics[width=0.45\linewidth]{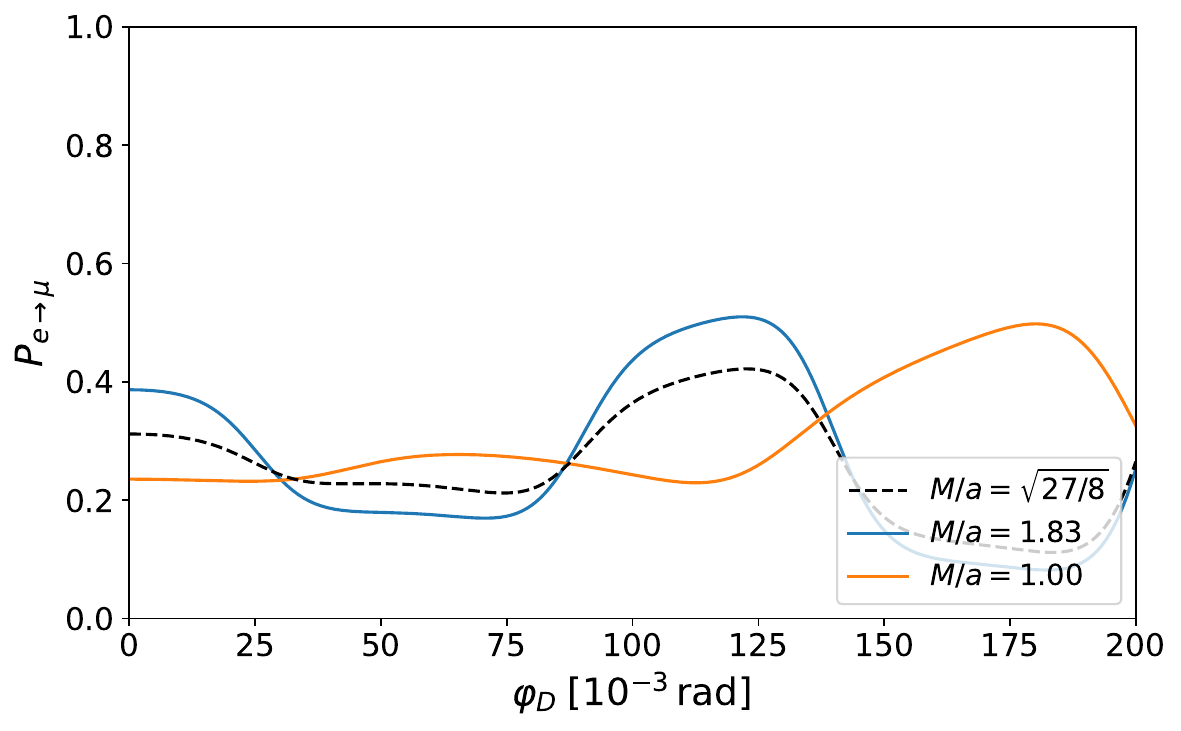}
        \caption{Two-flavor transition probability along three zero-winding paths in the critical and no-LR configurations of the dihole MP geometry. The upper panels show varying half-separation with \( a/M = \{ \sqrt{8/27},0.56,1.25\} \) at fixed \(M=10^5 M_\odot\). The lower panels show varying mass, \( M/a = \{ \sqrt{27/8},1.83, 1.00\} \) at fixed \(a=10^5 M_\odot \sqrt{8/27} \). The black dashed curve shows the critical configuration, which is the same in both comparisons. The source is fixed at \( (r_S,\varphi_S) = (10^3r_D, \pi) \), while the detector position varies over \( 0 \leq\varphi_D \leq 0.2 \) rad, at fixed \(r_D=10^6\)km. We set $E_0=3\times10^{10}\,\mathrm{eV}$, $|\Delta m^2|=10^{-3}\,\mathrm{eV}^2$, and $\theta=33.65^\circ$. The left panels correspond to $\Delta m^2>0$ ($m_1=0$), and the right panels correspond to $\Delta m^2<0$ ($m_2=0$).      
        }
        \label{fig:critical_overcritical}
    \end{figure}
    
    At \(\varphi_D = 0\), the two outer trajectories are accompanied by the radial \(b_p=0\) trajectory, which crosses the origin with no turning point. Away from alignment, each trajectory possesses one radial turning point satisfying the radial geodesic equation. Thus, the disappearance of LRs does not eliminate multiple connecting trajectories or their interference contributions to the neutrino transition probability. In \figref{fig:critical_overcritical} we show the probability for the critical case, represented by the black dashed curve, and two configurations with no LRs for each comparison considered. The upper panels correspond to a varying \(a\) with fixed \(M\), while the lower panels correspond to a varying \(M\) with fixed \(a\). In both cases, the configuration close to the critical value retains a similar oscillation pattern when compared to the critical case, although a shift in the amplitude is apparent. Further from criticality, both the heights and positions of the extrema change more substantially. Nevertheless, the critical and no-LR probabilities still remain oscillatory, with no clear monotonic increase or decrease in the probability for the spacetime configurations we studied. Thus, despite the differences in gravitational-lensing image formation \cite{MP_lensing_complete}, the critical and no-LR configurations considered here retain similar qualitative features in the zero-winding transition probability, while the heights and positions of the extrema depend on the spacetime parameters.      
    
    %
    %
    %
    %
    %
    %
    %
    %
    %
    %
    %%%%%%%%%%%%%%%%%%%%%%%%%%%%%%%%%%%%%%%%%%%%%%%%%%%%%%%%%%%%
    \section{Conclusion}\label{sec:conclusion}
    %%%%%%%%%%%%%%%%%%%%%%%%%%%%%%%%%%%%%%%%%%%%%%%%%%%%%%%%%%%%
    
    In this paper, we considered neutrino oscillations in the static equal-mass Majumdar-Papapetrou dihole geometry, restricting the propagation to the symmetry plane. Neutrino flavor states are understood as coherent superpositions of mass eigenstates that can propagate along multiple trajectories in curved spacetimes. We applied the null approximation in which the propagation paths are approximated by null geodesics shared by the mass eigenstates. We numerically evaluated the geodesic integrals and explored effects beyond the weak-lensing regime. Although our formulation is general, we restricted the numerical analysis to two neutrino flavors. In the model we used for neutrinos propagating in curved spacetimes, in addition to the interference due to the mass-squared difference, there is interference between different paths for both the same and different mass eigenstates. Consequently, the transition probability is sensitive to individual neutrino masses and can yield different probabilities for the two mass orderings.
    
    For the purely radial motion, we obtained a closed expression for the phase at leading order. For the non-radial case, that is, neutrino gravitational lensing, we considered zero-winding trajectories in three different cases based on the existence of LRs in the equatorial plane. For the configurations with two LRs, we restricted the analysis to the two primary zero-winding null geodesics connecting the source and detector, whose turning points lie outside the outer LR. For the critical and no-LR configurations, we numerically verified that three zero-winding trajectories connect the source and detector for the values we studied; thus, we considered these three trajectories.  
    
    For our probability studies, our setup fixes the source position, while the detector azimuthal coordinate is varied. For the weak-lensing configurations, the probability appears to increase as the black holes' individual masses increase, and no significant qualitative change could be observed by changing the black holes' half-separation. As strong-lensing configurations are considered, variations in the half-separation also modify the transition probability, suggesting that the dependence on the half-separation becomes more apparent in strong-lensing configurations. In the stronger-lensing configurations, however, the probability curves oscillate and intersect one another with no clear monotonic increase or decrease while changing the BH's individual masses or the separation between them. The oscillatory behavior is strongly sensitive to variations in the energy. Despite the significant difference in image formation for gravitational lensing due to the presence of LRs in the MP geometry \cite{MP_lensing_complete}, our transition probabilities remain oscillatory for all configurations we have studied. This analysis focused on the outer zero-winding trajectories for the two-LR configurations and three paths for the critical and no-LR configurations, and therefore cannot isolate the effect of the LR structure alone. Although the LR structure affects the available geodesics, differences in this structure do not appear to produce a distinct qualitative change in the transition probability in the configurations studied here. An interesting further extension would be to investigate the neutrino flavor oscillations near the merger of connecting trajectories in configurations without LRs.
    
    We emphasize that our studies have been restricted to the zero-winding trajectories. The formulation can be straightforwardly extended to include three flavors. The effects of extra windings around the gravitational center on the transition probability remain to be investigated; the corresponding trajectories can be studied using the already established techniques of strong gravitational lensing. However, including these trajectories poses conceptual and technical issues in treating quantum interference. Additional paths contribute to both the transition probability and the normalization -- including additional winding trajectories around the gravitational center would require incorporating their relative amplitudes consistently in both the propagated state and its normalization. Moreover, adding these paths introduces further interference terms between trajectories with different winding numbers. Extending the studies beyond the symmetry plane would allow an investigation of more general propagation geometries. Furthermore, in the general case, the null-geodesic motion in the MP geometry is not integrable \cite{chandrasekhar1989two,contopoulos1990periodic,contopoulos1991periodic,Dettmann:1994dj,Dettmann:1995ex}; the effects of chaotic trajectories on the neutrino flavor-transition probability are still an open question.    
    
    Studies of neutrino oscillations in gravitational fields often consider isolated black-hole backgrounds, exploiting spacetime symmetries and an ultrarelativistic expansion to relate the massive-particle constants of motion to those of a corresponding null geodesic. However, realistic astrophysical objects are often surrounded by accreting matter and electromagnetic fields. Extending these calculations to more general propagation geometries would broaden the understanding of such systems. Furthermore, constructing the theoretical foundation of neutrino oscillations in other compact and ultracompact objects could clarify how strong-field geometry influences neutrino propagation and flavor interference.  
    
    An important next step is to determine which lensing-induced flavor signatures survive wave-packet decoherence, finite source size, and detector energy resolution. The complementary capabilities of DUNE for supernova electron neutrinos \cite{DUNE:2021supernova} and IceCube-Gen2 for high-energy astrophysical neutrinos \cite{IceCubeGen2:2021} motivate extending this framework to realistic source spectra and three-flavor propagation. Such an analysis would establish whether spectral or flavor signatures remain observable after averaging, while the static MP dihole provides an exact benchmark for more realistic lens geometries.
    
    %
    %  
    %
    %%%%%%%%%%%%%%%%%%%%%%%%%%%%%%%%%%%%%%%%%%%%%%%%%%%%%%%%%%%%
    {\bf Acknowledgments}:  
    %%%%%%%%%%%%%%%%%%%%%%%%%%%%%%%%%%%%%%%%%%%%%%%%%%%%%%%%%%%%
    
    R.C. would like to thank PROEPD/UDESC for financial support. This research was supported in part by a fellowship by the Hanse-Wissenschaftskolleg -- Institute for Advanced Study (HWK).
    
    % \newpage
    %%%%%%%%%%%%%%%%%%%%%%%%%%%%%%%%%%%%%%%%%%%%%%%%%%%%%%%%%%%%
    
    %%%%%%%%%%%%%%%%%%%%%%%%%%%%%%%%%%%%%%%%%%%%%%%%%%%%%%%%%%%%%

    \end{document}